\documentclass[aps, showpacs, showkeys, nofootinbib, floatfix, superscriptaddress]{revtex4}

\usepackage{amsfonts}
\usepackage{amssymb}
\usepackage{amsmath}
\usepackage{graphicx}
\usepackage{float}
\usepackage{epstopdf}
\usepackage{amsmath}
\usepackage{makecell}
\usepackage{hyperref}
\usepackage{multirow}  
\usepackage{subcaption}

\usepackage{soulpos}      
\usepackage{xcolor}     
\usepackage{amsmath}       
\usepackage{hyperref}

\begin{document}

\title{Optical Appearance and Shadow of Kalb-Ramond Black Hole: Effects of Plasma and Accretion Models}
\author{Mou Xu}
\affiliation{Department of Physics, Liaoning Normal University, Dalian 116029, P. R. China}
\author{Ruonan Li}
\affiliation{Department of Physics, Liaoning Normal University, Dalian 116029, P. R. China}
\author{Jianbo Lu}
\email{lvjianbo819@163.com}
\affiliation{Department of Physics, Liaoning Normal University, Dalian 116029, P. R. China}
\author{Shining Yang}
\affiliation{Department of Physics, Liaoning Normal University, Dalian 116029, P. R. China}
\author{Shu-Min Wu}
\affiliation{Department of Physics, Liaoning Normal University, Dalian 116029, P. R. China}

\begin{abstract}
In this paper, we study the effect of the presence of plasma and different accretion models on the shadow and optical appearance of static spherically symmetric black holes containing the Kalb-Ramond field. We derive the motion equations for photons around the Kalb-Ramond black hole and constrain the Lorentz symmetry breaking parameters $\lambda$ and $\gamma$ using observational data released by the Event Horizon Telescope collaboration. The results indicate that, under the static spherical accretion model, as $\lambda$ or $\gamma$ increase, the peak value of the observed intensity for Kalb-Ramond black holes is enhanced and consistently exceeds that of the corresponding Schwarzschild black holes. In the presence of plasma, we find that as the plasma frequency increases, the photon sphere radius increases, whereas the black hole shadow radius decreases. In addition, compared to inhomogeneous plasma, the effect of homogeneous plasma on these features is more significant. Specifically, when the plasma is homogeneous, an increase in plasma frequency further enhances the observed intensity peaks. This suggests that the shadow of the Kalb-Ramond black hole is brighter due to the presence of plasma. Additionally, for the same Kalb-Ramond black hole model parameters and plasma frequency, the shadow of the Kalb-Ramond black hole in inhomogeneous plasma is larger than in the case of homogeneous plasma. Under the thin disk accretion model, an increase in the Lorentz symmetry breaking parameters decreases the observed intensity peak and increases the thickness of the photon ring and the lensed ring. The presence of plasma significantly alters the optical appearance of Kalb-Ramond black hole, providing a possible way to distinguish Kalb-Ramond black hole from Schwarzschild black hole.
\end{abstract}

\maketitle

\section{Introduction}

Black holes (BHs) are extremely compact celestial bodies predicted by General Relativity (GR), typically formed through the collapse of massive stars or other astrophysical processes. In recent years, scientists have conducted in-depth research on the existence of supermassive BHs by observing and analyzing astrophysical phenomena. In 2015, the Laser Interferometric Gravitational Wave Observatory successfully detected gravitational waves \cite{116.061102}, directly confirming the existence of BHs for the first time. Subsequently, the Event Horizon Telescope (EHT) collaboration released the first shadow image of the supermassive BH M87* in 2019 \cite{1906.11239,1906.11240,1906.11241}, and in 2022, they successfully observed the shadow structure of the SgrA* \cite{2311.08679,Event}. The center of these shadow images shows a dark region known as the "black hole shadow". The study of BH shadows was initiated by Synge and Luminet, who first provided the photon capture region for the Schwarzschild BH (SBH) \cite{Synge,Luminet}. Later, Bardeen analyzed the shadow of the Kerr BH and pointed out that its shadow is not a standard perfect circle \cite{Bardeen}. The BH shadow is considered the "fingerprint" of the BH, effectively reflecting the properties of the strong gravitational field around it. Therefore, it has become a key topic in the study of GR and modified gravity theories. Extensive literature on the study of BH shadows can be found in \cite{review1,review2,review3,chin1,chin2,chin3}.

Lorentz invariance is one of the cornerstones in the study of spacetime properties, assuming that physical laws are invariant in all inertial reference frames. However, some gravity theories suggest that Lorentz invariance may be broken in the ultraviolet regime, such as string theory \cite{PRD39,PRL63,PRL66}, loop quantum gravity \cite{0108061,9809038}, Horava-Lifshitz gravity  \cite{0411158}, and noncommutative field theory \cite{PRL87,0007031}. When the Lagrange density retains Lorentz invariant, but the system ground state breaks Lorentz symmetry, this phenomenon is known as spontaneous Lorentz symmetry breaking (LSB) \cite{9605088,0008252,1711.02273}. The study of spontaneous LSB is often carried out within the framework of Standard-Model extension \cite{0312310v2}. The bumblebee model is one of the simplest field-theory models, where a non-zero vacuum expectation value is obtained through its vector field, leading to the breaking of particle local Lorentz invariance \cite{ORD396831989,Phys.Rev. D 401886 (1989,VAKSS,Phys. Rev. D 74045001 (2006,theories,Phys.Rev.Lett.63224(1989}. This model has been widely studied in fields such as BHs \cite{1711.02273v1,1811.08503v2,s10052-020-7743-y}, wormhole \cite{1804.09911v2}, cosmology \cite{2407.13487v1,2411.18559v1}, and gravitational waves \cite{2207.14423v2,S0370269324003435-main}. In addition to vector fields, another rank-two antisymmetric tensor field may also lead to the local LSB of the particle, called the Kalb-Ramond (KR) field \cite{Phys.Rev.D101(2).024040(2020}. The KR field arises from the spectrum of the bosonic string theory \cite{physrevd.9.2273} and has received widespread attention in BH physics \cite{119f92711c9024a90254f8741f4bf5081,0210176v2,1611.06936v2}, cosmology \cite{2112.11945v2}, and brane world  \cite{1207.3152v3}. The literature \cite{2308.06613} derived (A)dS-Schwarzschild-like BH solutions in the KR theory, and subsequent references \cite{2106.14602v1,s10052-023-11231-5} investigated the quasinormal modes, photon sphere, and shadow properties of these BHs. \cite{s10052-024-13188} obtains charged BH solutions with the KR field background and investigates the corresponding thermodynamic properties. In this theory, the shadows of slowly rotating BHs have been studied in the literature \cite{2001.00460v2}. Moreover, traversable wormhole solutions can be constructed in KR theory \cite{2010.05298v1,s10052-022-10409-7}. In related research, we focus on the static spherically symmetric BH solution obtained in \cite{1911.10296}. This solution has been discussed in terms of geodesic structure \cite{10052.10619}, quasinormal modes \cite{2304.07761}, and gravitational lensing effects \cite{s10052-022-10619-z}. This paper aims to investigate the presence of plasma and the effect of different accretion models on the optical observational properties of static spherically symmetric BHs with the KR field background.

The BH images released by the EHT collaboration display its optical features, which primarily arise from the bending of light in the strong gravitational field of the BH. The dark region at the center of the image corresponds to the BH shadow, while the surrounding bright area is due to radiation from the accretion material \cite{2304.10015}. In astrophysics, BHs are usually surrounded by a large amount of accretion matter, whose distribution and physical state significantly influence the optical characteristics of the BH \cite{S0550321322003777}. Reference \cite{Image of a spherical black hole} first systematically studied BH images with thin accretion disks. It was further pointed out in \cite{1304.5691} that the shadow image can effectively distinguish between the SBH and the static wormhole. Additionally, spherical accretion models have been widely applied in the study of BH images \cite{Falcke_2000_ApJ_528_L13,Narayan_2019_ApJL_885_L33}. Observations from the EHT collaboration also suggest that plasma and magnetic fields are usually present around astrophysical BHs \cite{2212.12949}. The presence of plasma not only affects the trajectory of photons but also significantly impacts the optical appearance of the BH. In a vacuum, the trajectory of photons follows geodesics, whereas in a dense charged medium (such as plasma near a BH), the trajectory of photons will differ substantially \cite{1507.08545,1905.06615}. Recently, literature \cite{1507.04217v2} studied the impact of plasma on the shadow of spherically symmetric BHs, and \cite{1702.08768,1507.08131} has further explored the effect of the plasma on the shadow shape of rotating BHs. It is shown that the shadow of a Kerr BH shrinks and tends to become more circular in a denser plasma environment \cite{1807.06268}. Furthermore, in various modified gravity theory frameworks, the influence of plasma on gravitational lensing effects and BH shadows has been widely discussed \cite{lrr-2004-9,1702.08768v2,1507.04217v2}.
This paper will analyze the effects of the static spherical accretion model without plasma, the static spherical accretion model with the presence of plasma around the KRBH, and the thin disk accretion model without plasma on the observational characteristics of the KRBH, respectively.
 
This paper is organized as follows: In section 2, we briefly introduce the static spherically symmetric BH solutions in the framework of KR theory. Using the geodesic equations, we derive the motion equations for photons around the KRBH. Additionally, we constrain the LSB parameters $\gamma$ and $\lambda$ using observational data released by the EHT. The influence of the static spherical accretion model without plasma on the observational properties of the KRBH is examined in section 3. Section 4 studies the BH images under the static spherical accretion model in the presence of plasma around the KRBH. Further, we analyze the behavior of the photon sphere radius, shadow radius, and optical appearance of the KRBH under homogeneous and inhomogeneous plasma. Section 5 explores the effects of three thin accretion disk models on the optical appearance of the KRBH. We use geometric units with $G=c=M=1$ in the calculations throughout the paper.

\section{Static spherically symmetric black hole solutions with Kalb-Ramond field}

The KR field is a tensor field arising from the spectrum of the bosonic string theory, represented by a 2-form tensor $B_{\mu\nu}$, which satisfies the condition $B_{\mu\nu} = -B_{\nu\mu}$. The self-interaction potential $V$ depends on $B_{\mu\nu}B^{\mu\nu}$ to ensure the theory invariant under local Lorentz transformations of the observer. Suppose the potential has the form $V= V(B_{\mu\nu}B^{\mu\nu}\pm \bar{b}^2)$, where $\bar{b}^2$ is a positive constant, and this potential corresponds to a non-vanishing vacuum expectation value $\langle B_{\mu \nu}\rangle=\bar{b}_{\mu \nu}$. Due to the non-minimal coupling of the KR field with the gravitational field, the non-vanishing vacuum expectation value $\bar{b}_{\mu \nu}$ spontaneously breaks the local Lorentz symmetry of the particles  \cite{1911.10296}. The form of the action is as follows \cite{Phys. Rev. D 81}:
\begin{equation}
	\begin{aligned}
	\mathcal{I}_{K R}=\int \tilde{g} d^{4} x\left[\frac{R}{2 \kappa}-\frac{1}{12} H_{\lambda\mu \nu} H^{\lambda\mu \nu}-V\left(B_{\mu \nu} B^{\mu \nu} \pm \bar{b}_{\mu \nu} \bar{b}^{\mu \nu}\right)+\frac{1}{2 \kappa}\left(\xi_{2} B^{\lambda \mu} B^{\mu}_{\nu} R_{\lambda\nu}+\xi_{3} B^{\mu \nu} B_{\mu \nu} R\right)\right],
	\label{eq21}
	\end{aligned}
\end{equation}
where $\xi_{2}$ and $\xi_{3}$ are non-minimal coupling constants, $\tilde{g}$ is the metric determinant, $\kappa={8\pi G}$ is the gravitational coupling constant, and $R$ is the Ricci scalar. $H_{\lambda \mu \nu} \equiv \partial_{[\lambda} B_{\mu \nu]}$ is the KR field strength tensor \cite{0912.4852v1}. By varying the action (\ref{eq21}) with respect to the metric $ g_{\mu \nu} $, the modified Einstein field equation can be derived as:
\begin{equation}
	R_{\mu \nu}-\frac{1}{2} R g_{\mu \nu}=\kappa T_{\mu \nu}^{\xi_{2}},
	\label{eq22}
\end{equation}
where
\begin{equation}
	\begin{aligned}
	T_{\mu\nu}^{\xi_{2}}=&\frac{\xi_{2}}{\kappa}\left[\frac{1}{2} g_{\mu \nu} B^{\alpha \gamma} B_{\gamma}^{\beta} R_{\alpha \beta}-R_{\mu}^{\alpha} B_{\mu}^{\beta} R_{\alpha \beta}-B^{\alpha \beta} B_{\mu \beta} R_{\nu \beta}-B^{\alpha \beta} B_{\nu \beta} R_{\mu \alpha}+\frac{1}{2} D_{\alpha} D_{\mu}\left(B_{\nu \beta} B^{\alpha \beta}\right)\right. \\
	&\left.+\frac{1}{2} D_{\alpha} D_{\nu}\left(B_{\mu \beta} B^{\alpha \beta}\right)-\frac{1}{2} D^{2}\left(B_{\mu}^{\alpha} B_{\alpha \nu}\right)-\frac{1}{2} g_{\mu \nu} D_{\alpha} D_{\beta}\left(B^{\alpha \gamma} B_{\gamma}^{\beta}\right)\right]. 
	\end{aligned}
	\label{eq24}
\end{equation}

In the static spherically symmetric spacetime, the line element can be expressed as:
\begin{equation}
	ds^{2}=-A(r)dt^{2}+\frac{1}{A(r)}dr^{2}+r^{2}d\theta^{2}+r^{2}\sin^{2}\theta d\phi^{2}.\label{eq25}
\end{equation}
The literature \cite{1911.10296} derives the Schwarzschild-like BH solution with the KR field background as follows:
\begin{equation}
A(r)=1-\frac{2 G M}{r}+\frac{\gamma}{r^{2 / \lambda}},
\label{eq26}
\end{equation}
where $\lambda$ and $\gamma$ are parameters related to spontaneous LSB. $\lambda$ is defined as the product of the norm of the KR vacuum expectation value, $\bar{b}^{2}$, and the non-minimal coupling constant $\xi_{2}$, i.e., $\lambda=\bar{b}^{2} \xi_{2}$. Since the Lorentz violating effects on the gravitational field are expected to be very small, the coupling constant $\xi_2$ is assumed to be small. On the other hand, Lorentz violating is expected to originate near the Planck scale, and thus the $\bar{b}^2$ is expected to be of similar magnitude. This configuration — featuring a large $\bar{b}^2$ and a small coupling constant $\xi_{2}$ — is naturally realized in frameworks involving spontaneous LSB \cite{1911.10296}. $\gamma$ is an integration constant. Unlike the BH mass, it has no Newtonian analogue to determine $\gamma$. When $\lambda \rightarrow 0$ or $\gamma=0$, the BH solution (\ref{eq26}) reduces to the SBH solution. It is worth noting that in the special case of $\lambda=1$, the Eq.(\ref{eq26}) owns a similar form to the Reissner-Nordström black hole (RNBH) solution. However, the pseudo-electric field within the framework of the KR gravity is a constant, unlike the RN case. This indicates that $\gamma$ cannot be interpreted as an electric charge, but rather as a Lorentz violating "hair" associated with the BH \cite{1911.10296}.
By setting $A(r) \Big|_{r=r_e} = 0$, we obtain the equation that the event horizon radius $r_e$ of the KRBH satisfies:
\begin{equation}
	r_e^{\frac{2+\lambda}{\lambda}} - 2Mr_e^{\frac{2}{\lambda}} + \gamma r_e = 0.
	\label{eqre}
\end{equation}

\subsection{The Geodesic Structure of Photons for KRBH}

Next we will focus on the motion of photons around KRBH. In the framework of KR gravity, the Hamiltonian of photons around a static spherically symmetric BH can be written as:
\begin{equation}
	\mathcal{H}=\frac{1}{2} g^{\mu \nu} p_{\mu} p_{\nu}=\frac{1}{2}\left[-A(r) \dot{t}^{2}+\frac{1}{A(r)} \dot{r}^{2}+r^{2}\left(\dot{\theta}^{2}+\sin^{2} \theta \dot{\phi}^{2}\right)\right],
	\label{eq27}
\end{equation}
where $p_{\mu}$ represents the four velocity of the photon, and ''dot'' denotes the derivative of the general coordinate with respect to the affine parameter $s$. Since the metric coefficients are not explicitly depend on time $t$ and azimuthal angle $\phi$. We can derive two corresponding conserved quantities along with the motion of photons:
\begin{equation}
E=-\frac{\partial \mathcal{H}}{\partial p_t}=\left(1-\frac{2 G M}{r}+\frac{\gamma}{r^{2 / \lambda}}\right) \dot{t},
\label{eq28}
\end{equation}
\begin{equation}
L=\frac{\partial \mathcal{H}}{\partial p_\phi}=r^{2} \dot{\phi}.
\label{eq29}
\end{equation}
$E$ and $L$ represent the energy and angular momentum of the photon, respectively. Furthermore, without loss of generality, we consider the motion of the photon in the equatorial plane, i.e., $\theta=\frac{\pi}{2}$, $\dot{\theta}=0$. By defining the impact parameter $b \equiv L/E$ and redefining the affine parameter $s$ as $s/L$, the equations of motion for photons around the KRBH are given by:
\begin{equation}
	\dot{t} =\frac{1}{b(1-\frac{2 M}{r}+\frac{\gamma}{r^{2 / \lambda}})},
	\label{eq210}
\end{equation}
\begin{equation}
	\dot{r}^2=\frac{1}{b^{2}}-\frac{1-\frac{2 M}{r}+\frac{\gamma}{r^{2 / \lambda}}}{r^{2}},
	\label{eq211}
\end{equation}
\begin{equation}
	\dot{\phi} =\frac{1}{r^{2}}.
	\label{eq212}
\end{equation}
In addition, the Eq. (\ref{eq211}) can be rewritten as $\dot{r}^2=\frac{1}{b^{2}}-V_{eff}$. The $V_{eff}$ is the effective potential of photons around the KRBH:
\begin{equation}
	V_{e f f}\equiv\frac{1-\frac{2 M}{r}+\frac{\gamma}{r^{2/ \lambda}}}{r^{2}}.
	\label{eq213}
\end{equation}
\begin{figure}[H]
\centering
\includegraphics[width=8cm]{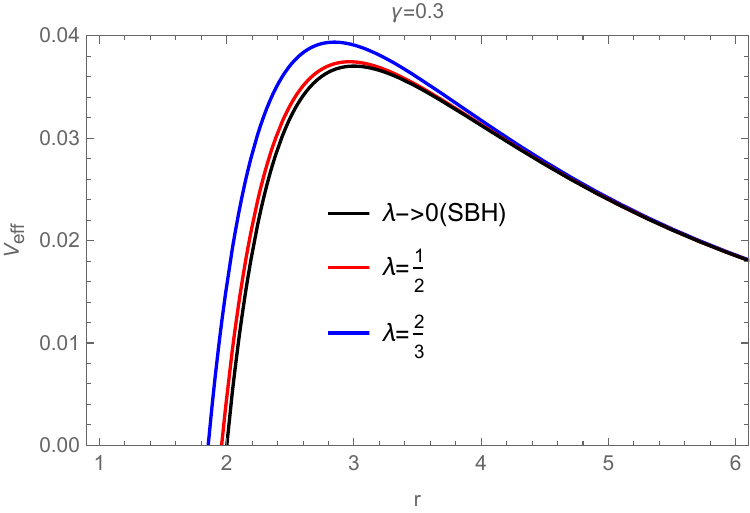}
\includegraphics[width=8cm]{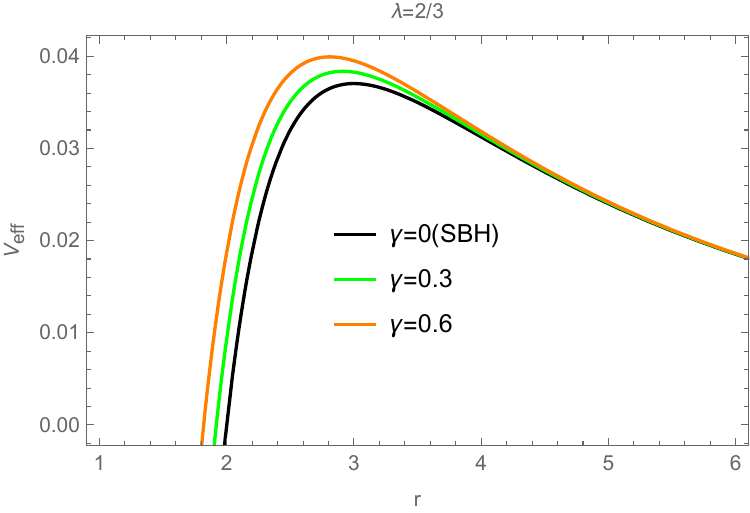}
\caption{The variation of the effective potential $V_{eff}$ of photons around KRBH with respect to the radial coordinate $r$. The left panel corresponds to $\gamma=0.3$, and the right panel corresponds to $\lambda= 2/3$.}
\label{fig21}
\end{figure}
\noindent
The sign of the second derivative of the effective potential can be used to determine the stability of photon orbits \cite{2412.00796}, i.e., when $\frac{d^{2} V_{e f f}}{d r^{2}}>0$, the orbit is stable, and when $\frac{d^{2} V_{e f f}}{d r^{2}}<0$, the orbit is unstable. Based on Eq. (\ref{eq213}), we plot the variation of the effective potential $V_{eff}$ of photons around KRBH with respect to the radial coordinate $r$ in Fig. \ref{fig21}. From the figure, we can observe that the peak of the effective potential for photons increases as the LSB parameters $\gamma$ and $\lambda$ increase. At the event horizon, the effective potential is zero, and as the radial coordinate increases, the effective potential reaches a peak at $r=r_{ph}$ and subsequently decreases as the radial coordinate increases. The $r_{ph}$ is the photon sphere radius of the KRBH, and photons on this orbit will travel around the BH infinite times without perturbation. According to the conditions $V_{e f f}(r_{ph})=\frac{1}{b_{ph}^{2}}$ and $\frac{dV_{eff}}{dr} \big|_{r=r_{ph}}=0$, we can derive that the photon sphere radius $r_{p h}$ of the KRBH satisfies the expression:
\begin{equation}
	-3 M r_{p h}^{2 / \lambda} \lambda+r_{p h}^{(2+\lambda) / \lambda} \lambda+r_{p h} \gamma(1+\lambda)=0,
	\label{eq214}
\end{equation}
as well as the expression for the critical impact parameter $b_{p h}$:
\begin{equation}
	b_{ph} = \frac{r_{ph}^{\frac{1}{2}(3 + \frac{2}{\lambda})}}{\sqrt{r_{ph}^{1 + 2/\lambda} - 2M r_{ph}^{2/\lambda} + r_{ph} \gamma}}.\label{eq215}
\end{equation}
For an observer at infinity, $b_{ph}$ corresponds to the shadow radius of the BH $R_{s h}$. Based on Eqs. (\ref{eqre}), (\ref{eq214}), and (\ref{eq215}), we present the specific values for the event horizon radius $r_e$, the photon sphere radius $r_{p h}$ and shadow radius $R_{s h}$ for different values of $\gamma$ and $\lambda$ in Table \ref{tab1} and Table \ref{tab11}, respectively. From the tables, it can be seen that as the parameters $\gamma$ or $\lambda$ increase, the event horizon, photon sphere, and BH shadow of the KRBH exhibit an inward contraction trend, and all are smaller than the corresponding quantities for the SBH. Given the critical role of the innermost stable circular orbit (ISCO) in BH physics and its importance for the subsequent analysis, we present the expression for the ISCO radius $r_{ISCO}$ of the KRBH \cite{2112.11227, huyapeng}:
\begin{equation}
r_{ISCO}=\frac{3A(r_{ISCO})A'(r_{ISCO})}{2A'(r_{ISCO})^2-A(r_{ISCO})A''(r_{ISCO})}.
\label{isco}
\end{equation}
Based on this formula, the values of the ISCO radius for different $\gamma$ and $\lambda$ are also listed in Table \ref{tab1} and Table \ref{tab11}, respectively.
\begin{table}[H]
	\centering
	\caption{Event horizon radius, photon sphere radius, shadow radius and ISCO radius of KRBH for different $\gamma$ values when $\lambda=2/3$.}
	\begin{tabular}{|c|c|c|c|c|}
		\hline
		&$\gamma=0$ (SBH)  &$\gamma=0.3$    &$\gamma=0.6$  &$\gamma=0.9$ \\
		\hline
		$r_{e}$&2.000  &1.918   &1.819  &1.682 \\
		\hline
		$r_{ph}$&3.000  &2.912    &2.810  &2.689\\
		\hline        
		$R_{s h}$&5.196  &5.101   &5.004  &4.889 \\
		\hline   
		$r_{ISCO}$&6.000  &5.838  &5.656  &5.457 \\
		\hline       
	\end{tabular}
	\label{tab1}
\end{table}

\begin{table}[H]
	\centering
	\caption{Event horizon radius, photon sphere radius, shadow radius and ISCO radius of KRBH for different $\lambda$ values when $\gamma=0.3$.}
	\begin{tabular}{|c|c|c|c|c|}
		\hline
		&$\lambda \to 0$ (SBH)    &$\lambda=\frac{1}{2}$  &$\lambda=\frac{2}{3}$  &$\lambda=1$ \\
		\hline       
		$r_{e}$&2.000    &1.961  &1.918  &1.837  \\
		\hline      
		$r_{ph}$&3.000    &2.965  &2.912  &2.785 \\
		\hline      
		$R_{s h}$  &5.196  &5.167  &5.101  &4.919\\
		\hline  
		$r_{ISCO}$&6.000  &5.953  &5.835  &5.230 \\
		\hline 
	\end{tabular}
	\label{tab11}
\end{table}

\subsection{Constraints on Model Parameters of KRBH Using EHT Data}

In order to test the accuracy of gravity theories in the strong gravitational field of BH, it is necessary to use EHT data to constrain BH model parameters \cite{review2}. For example, in \cite{2406.07300}, EHT observational data were used to constrain the parameters of magnetically charged regular BHs, and in \cite{Afrin2022.93251}, the parameters of rotating Horndeski BHs were constrained. In the following, we use the shadow angular diameter data of SgrA* released by EHT to constrain the LSB parameters $\gamma$ and $\lambda$ of the static spherically symmetric KRBH. 

\begin{figure}[H]
	\centering
	\includegraphics[width=8cm]{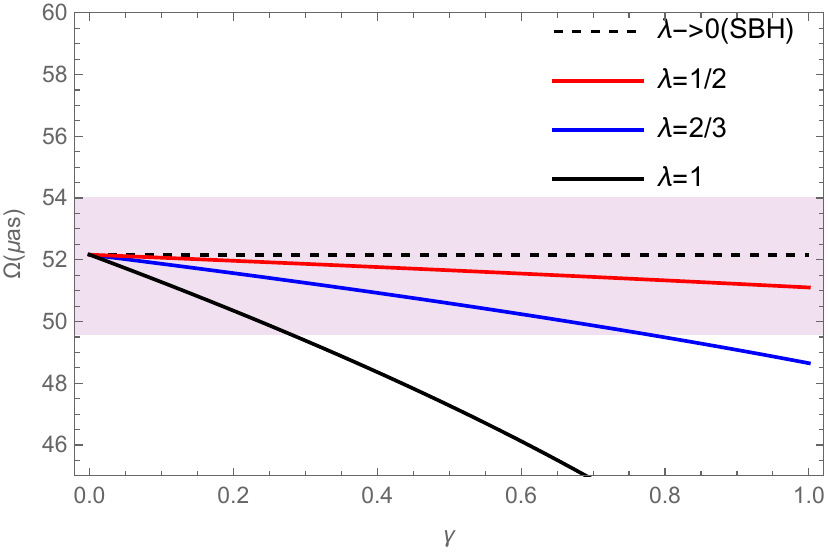}
	\includegraphics[width=8cm]{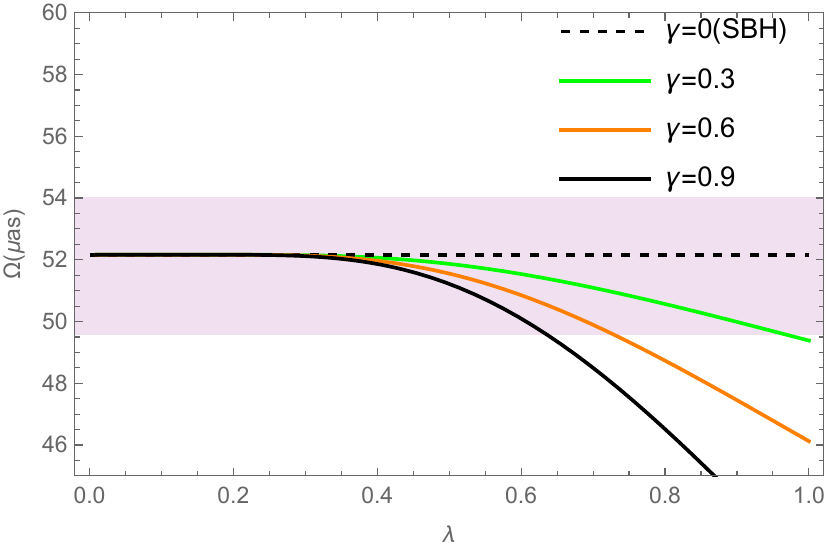}
	\caption{Variation of the shadow angular diameter of the static spherically symmetric KRBH relative to parameters $\gamma$ and $\lambda$, along with the constraints on parameters $\gamma$ and $\lambda$ from shadow angular diameter of the SgrA*.}
	\label{fig22}
\end{figure}

For a distant observer, the angular diameter $\Omega$ of the BH shadow can be defined as \cite{review2}:
\begin{equation}
	\Omega=2 \frac{b_{p h}}{D},\label{eq216}
\end{equation}
where $D$ represents the distance between the BH and the observer. Further, the above equation can be written as:
\begin{equation}
	\left(\frac{\Omega}{\mu a s}\right)=\left(\frac{6.191165 \times 10^{-8}}{\pi} \frac{\widetilde{\gamma}}{D / M p c}\right)\left(\frac{b_{p h}}{M}\right) \\,\label{eq217}
\end{equation}
where $\widetilde{\gamma}$ is the mass ratio of the BH to the Sun. According to the data released by the EHT, two parameters for SgrA* are $\widetilde{\gamma}= 4.3 \times 10^6$ and $D=8.27kpc$ \cite{Event,2205.07787.can22,2205.07787.can18}. Using Eqs. (\ref{eq215}) and (\ref{eq217}), Fig. \ref{fig22} displays the variation of the shadow angular diameter of the static spherically symmetric BH with the KR field background relative to the parameters, along with the constraints on parameters $\gamma$ and $\lambda$ based on the shadow angular diameter data of SgrA* ($51.8 \pm 2.3 \, \mu\text{as}$). From the figures, we observe that the shadow angular diameter of the KRBH decreases as the parameters $\gamma$ and $\lambda$ increase. In Table \ref{tab23} and Table \ref{tab24}, we show the constraint results on the other parameter when $\lambda$ and $\gamma$ take different values, respectively. 

In addition, one can notice that some theoretical analyses and observational data have already placed constraints on the LSB parameters. For instance, the weak energy condition is satisfied for $0 \leq  \lambda  \leq 2$ and $\gamma >0$ \cite{1911.10296}. Shadow radius from the EHT have set upper bounds on $\lambda$, with $\lambda \lesssim 1.2$ $(1 \sigma )$ and $\lambda \lesssim 1.5$ $(2 \sigma )$ for $\gamma =1/2$ \cite{2205.07787}. Evidently, our constraints fall within the ranges established in the aforementioned results.
\begin{table}[H]
\centering
\caption{Constraints on $\gamma$ from shadow angular diameter of the SgrA* for different values of the LSB parameter $\lambda$.}
\begin{tabular}{|c|c|c|c|}
	\hline
	$\lambda$&$1/2$    &$2/3$ &$1$   \\
	\hline       
	Constraints on $\gamma$&$\gamma \lesssim  2.149$ &$\gamma \lesssim  0.797$  &$\gamma \lesssim  0.289$  \\
	\hline      
\end{tabular}
\label{tab23}
\end{table}

\begin{table}[H]
\centering
\caption{Constraints on $\lambda$ from shadow angular diameter of the SgrA* for different values of the LSB parameter $\gamma$.}
\begin{tabular}{|c|c|c|c|}
	\hline
	$\gamma$& 0.3    &$0.6$ &$0.9$   \\
	\hline       
	Constraints on $\lambda$&$\lambda \lesssim  0.981$    &$\lambda \lesssim  0.736$  &$\lambda \lesssim  0.641$  \\
	\hline      
\end{tabular}
\label{tab24}
\end{table}

\section{Shadows and Photon Rings of Kalb-Ramond Black Holes in Static Spherical Accretion Without Plasma}

Due to the intrinsic properties of BHs, the radiation we observe does not come from the BH, but mainly from the accretion disk surrounding it \cite{Irr 2013-1}. The shadow and photon ring features of BHs depend not only on the spacetime structure but also on the characteristics of the accretion disk. Photons may orbit the BH multiple times before escaping, forming several bright rings. By studying these optical phenomena, we can test the predictions of gravity theory in strong gravitational fields and deepen our understanding of the nature of spacetime.

In this section, we assume that in the absence of plasma, the accretion material around the BH is static. When the angular momentum of the accretion material is zero, spherical accretion forms around the BH. Next, we will investigate the optical properties of KRBHs surrounded by optically and geometrically thin static spherically accretion flow. For an observer located at infinity, the observed specific intensity can be written as:
\begin{equation}
	I_{obs} = \int g^3 j(\nu_{em}) dl_{prop},
	\label{eq432}
\end{equation}
where $g=\sqrt{A(r)}$ is the redshift factor, $\nu_{em}$ is the emit photon frequency, and $j(\nu_{em})$ is the emission rate per unit volume \cite{1304.5691}. For monochromatic emission, we assume $j(\nu_{em}) \propto \frac{1}{r^2}$. $dl_{prop}$ represents the infinitesimal proper length, which in the KR spacetime takes the form:
\begin{equation}
	dl_{prop} = \sqrt{\frac{1}{1-\frac{2 M}{r}+\frac{\gamma}{r^{2 / \lambda}}} + r^2 \left(\frac{d\phi}{dr}\right)^2} dr,
	\label{eq433}
\end{equation}
where $\frac{d \phi}{dr}$ can be obtained by solving Eqs. (\ref{eq211}) and (\ref{eq212}). Using Eqs. (\ref{eq432}-\ref{eq433}), the total intensity observed by a distant static observer is:
\begin{equation}
	I_{obs} = \int \frac{(1-\frac{2 M}{r}+\frac{\gamma}{r^{2 / \lambda}})^{3/2}}{r^2} \sqrt{\frac{1}{1-\frac{2 M}{r}+\frac{\gamma}{r^{2 / \lambda}}} + \frac{b^2}{r^2 - b^2 (1-\frac{2 M}{r}+\frac{\gamma}{r^{2 / \lambda}})}} dr.
	\label{eq435}
\end{equation}
\begin{figure}[H]
	\centering
	\includegraphics[width=8cm]{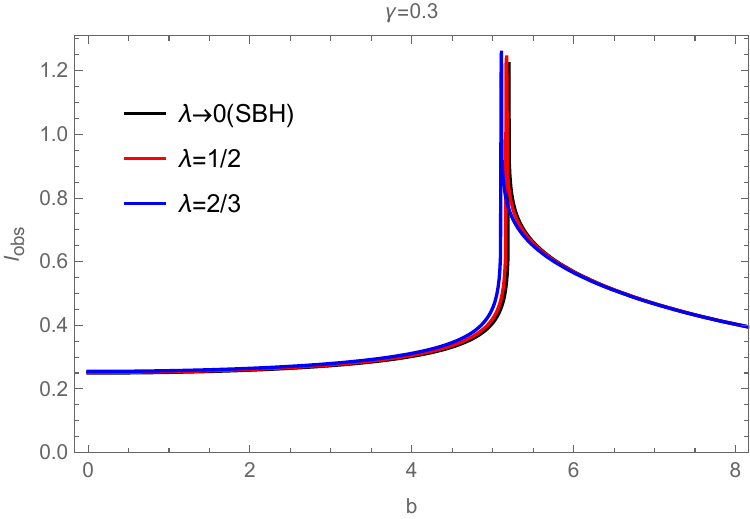}
	\includegraphics[width=8cm]{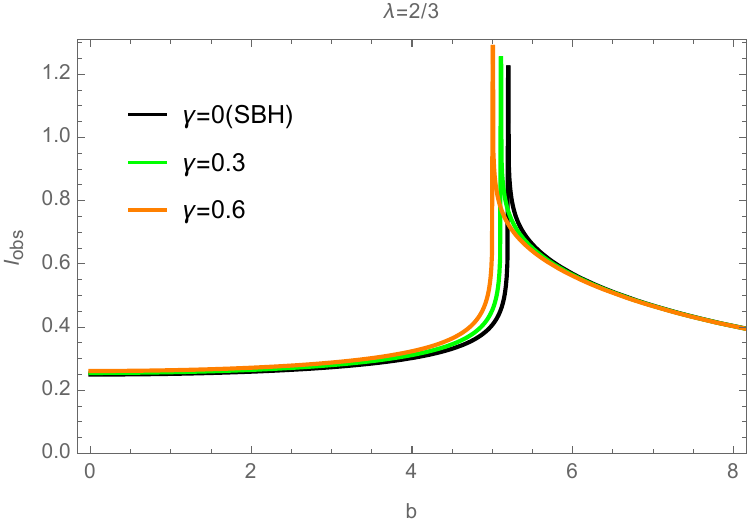}
	\caption{Variation of the observed intensity $I_{obs}$ with the impact parameter $b$ around the KRBH under static spherical accretion. The left panel corresponds to $\gamma=0.3$, and the right panel corresponds to $\lambda=2/3$.}
	\label{fig29}
\end{figure}
\noindent According to Eq. (\ref{eq435}), in Fig. \ref{fig29} we plot the variation of the observed intensity of KRBH under the static spherical accretion model with respect to the impact parameter $b$. The left panel shows the effect of the LSB parameter $\lambda$ on the observed intensity when $\gamma=0.3$. The right panel illustrates the impact of the parameter $\gamma$ on the observed intensity when $\lambda=2/3$. From the figure, we can observe that as the value of $b$ increases, the total observed intensity gradually increases, reaching a maximum value at the critical impact parameter $b_{ph}$ and then decreases. Notably, the increase in the LSB parameters $\lambda$ and $\gamma$ both lead to an increase in the total observed intensity. In other words, compared to the SBH, the KRBH exhibits higher observed intensity, which can serve as a distinguishing feature between the two types of BHs. Furthermore, in Fig. \ref{fig210} and Fig. \ref{fig2101}, we show the effects of the LSB parameter $\gamma$ on the optical appearance of KRBH when $\lambda=2/3$, as well as the effects of $\lambda$ on the optical appearance of KRBH when $\gamma=0.3$. From the figures, we observe that the bright ring outside the BH shadow corresponds to the photon sphere, where the observed intensity is maximum. Additionally, due to the static spherically symmetric accretion flow, a tiny fraction of the radiating gas behind the BH can escape to infinity, and thus the observed intensity in the shadow is not zero.
\begin{figure}[H]
	\centering
	\includegraphics[width=5cm]{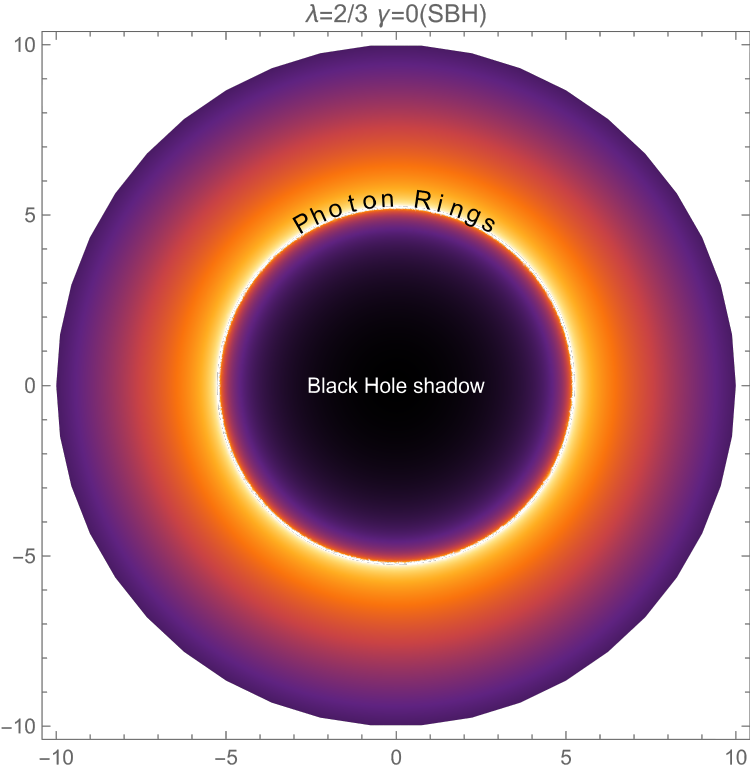}
	\includegraphics[width=5cm]{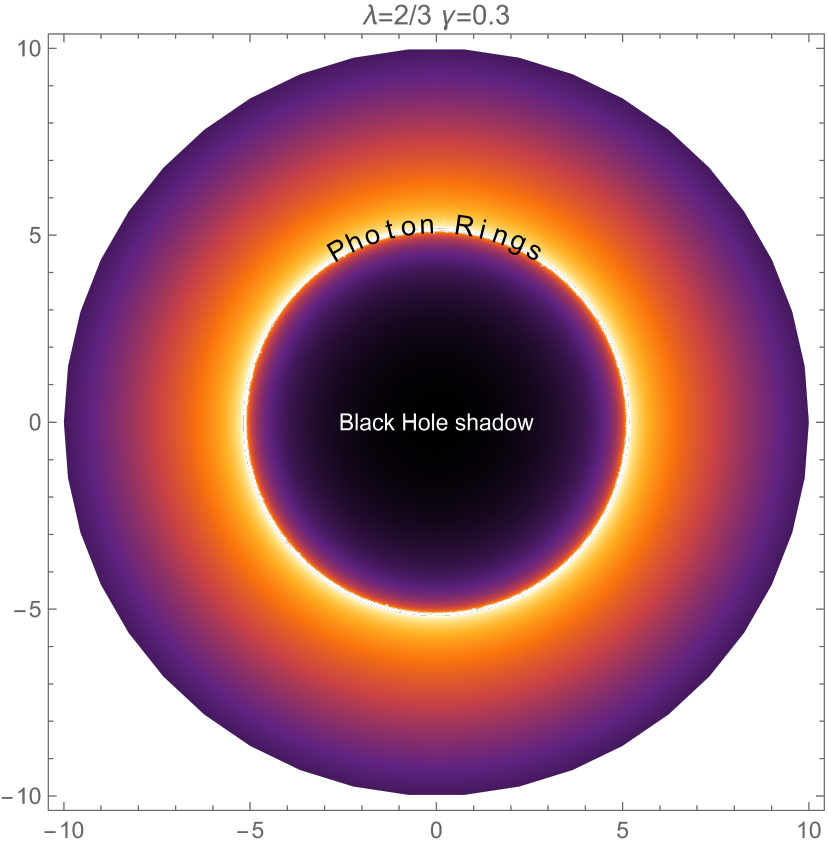}
	\includegraphics[width=5cm]{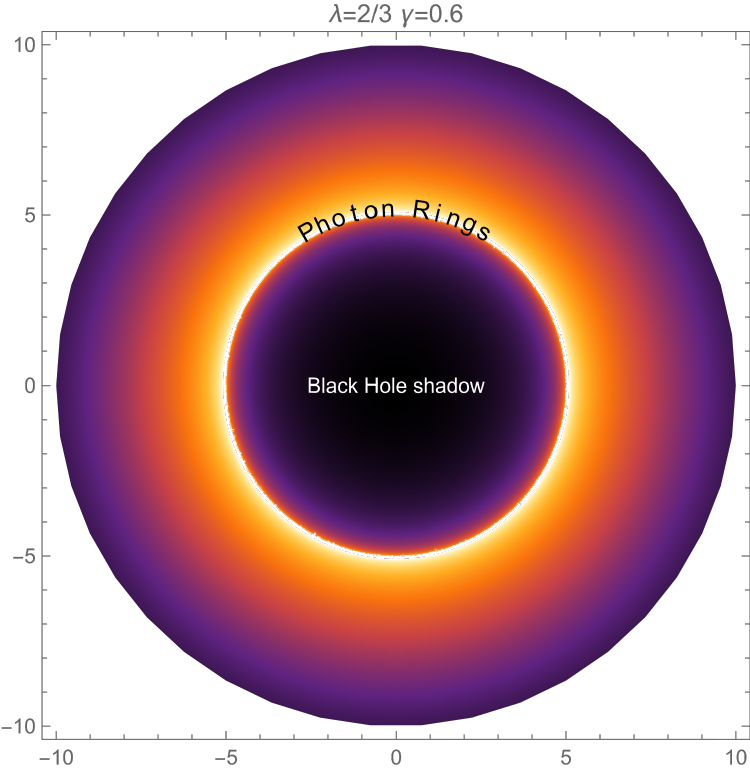}
	\caption{When parameter $\lambda=2/3$, two dimensional images of the shadow and photon ring of the KRBHs under static spherical accretion with different values of the LSB parameter $\gamma$.}
	\label{fig210}
\end{figure}
\begin{figure}[H]
	\centering
	\includegraphics[width=5cm]{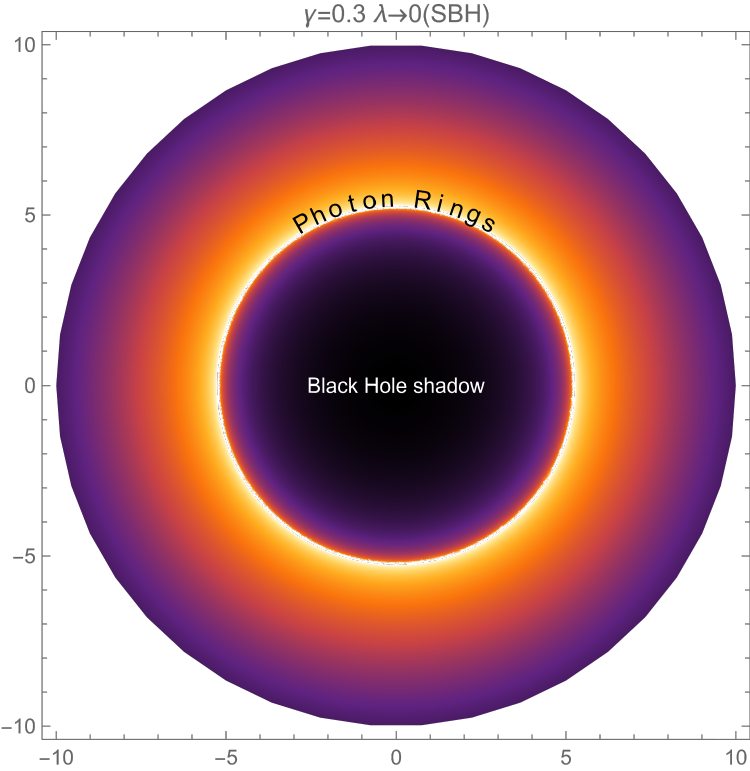}
	\includegraphics[width=5cm]{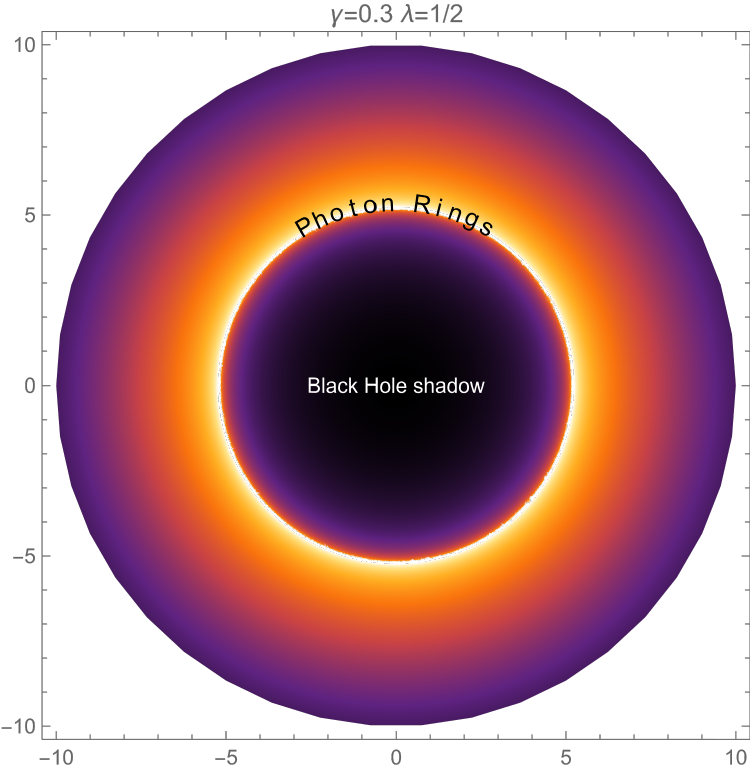}
	\includegraphics[width=5cm]{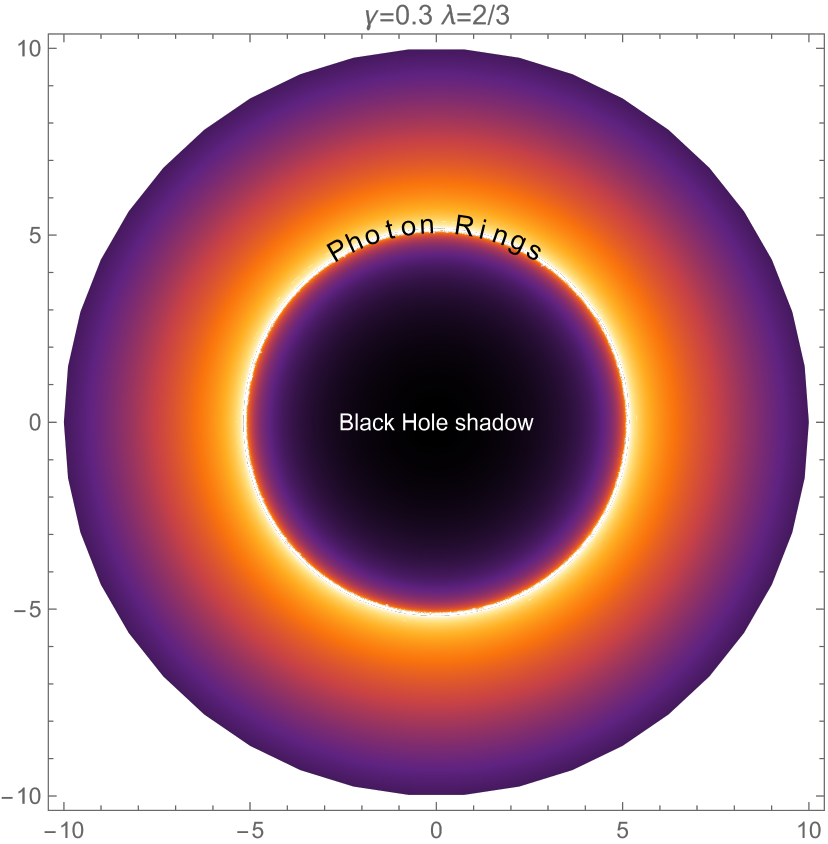}
	\caption{When parameter $\gamma=0.3$, two dimensional images of the shadow and photon ring of the KRBHs under static spherical accretion with different values of the LSB parameter $\lambda$.}
	\label{fig2101}
\end{figure}

\section{The Influence of Plasma on the Optical Appearance of Kalb-Ramond Black Holes in Spherical Accretion}

Current astrophysical observations indicate that supermassive BHs are typically surrounded by plasma and magnetic field environments. The image of M87* released by EHT and multi-wavelength observations, show that the BH is surrounded by a relativistically hot, magnetized plasma \cite{2105.01173}. The EHT collaboration further estimated the mass accretion rate of the central BH in M87* to be $(3-20) \times 10^{-4}$ solar masses per year \cite{2312.10678}. This represents the first observational result confirmation of the presence of a plasma structure in the vicinity of a supermassive BH. Furthermore, some studies have shown that the plasma number density around Sgr A* and M87* is currently estimated to be $N \sim 10^{10}cm^{-3}$ \cite{2206.04430}.
Perlick et al. have extensively studied the dynamical models of non-magnetized plasma around spherically symmetric and axisymmetric BHs \cite{1702.08768,1507.04217v2,review2}. Subsequently, the impact of plasma on the BH shadows in the context of various modified gravity theories has garnered significant attention \cite{2209.01652,Eur. Phys. J. C 82 771 (2022,2106.07601,2110.11704}. In addition, gravitational lensing effects in plasmas have been extensively studied \cite{Eur. Phys. J. C 82 659 (2022,Eur. Phys. J. Plus 137 634 (2022,Phys.Rev. D 104. 084015 (2021,2201.09879}. Whether plasma exists around a BH and whether its medium is homogeneous both influence the motion of photons around the BH \cite{1507.08545,1905.06615}. In the previous section, we studied the BH shadow and optical appearance of KRBHs in the absence of plasma. Here, we now discuss the characteristics of the KRBH optical appearance in the presence of plasma.

\subsection{Photon Motion in Plasma Surrounding Kalb-Ramond Black Hole}
\begin{figure}[H]
	\centering
	\includegraphics[width=8cm]{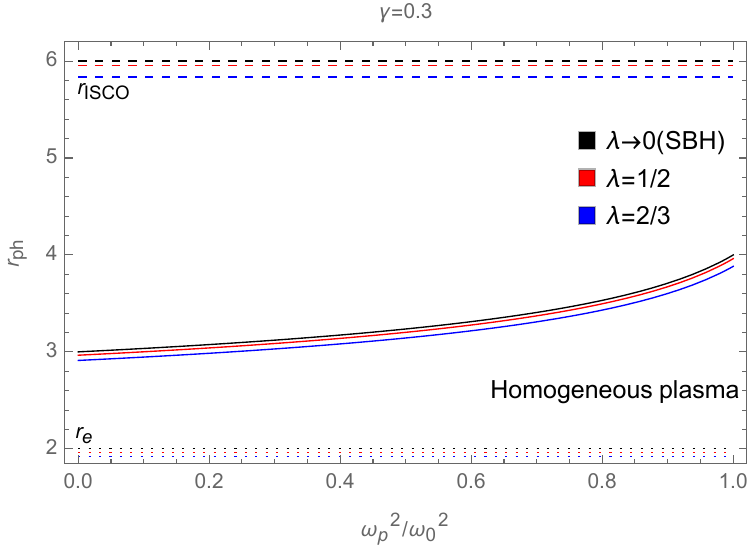}
	\includegraphics[width=8cm]{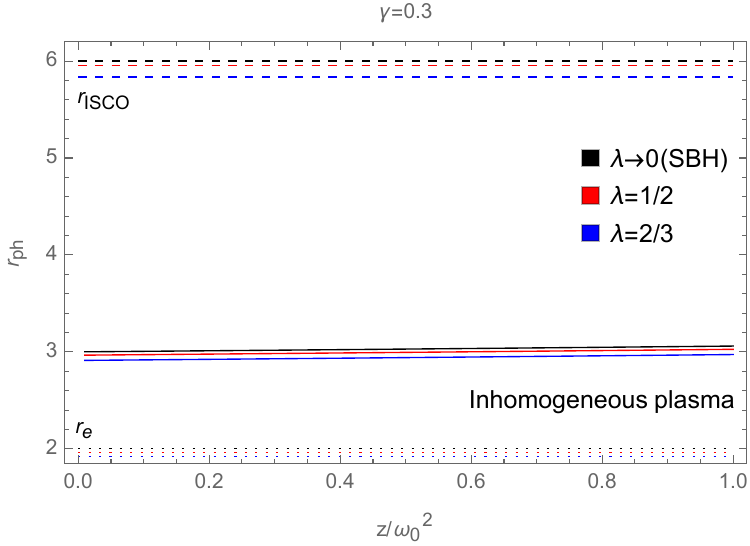}
	\includegraphics[width=8cm]{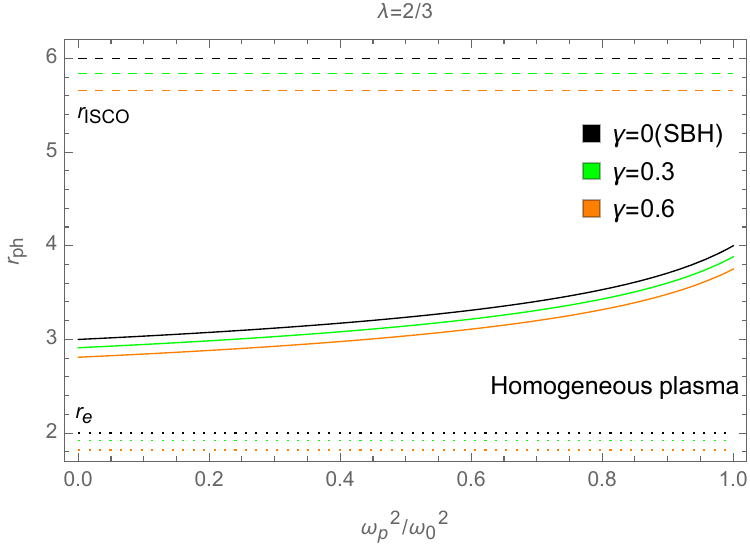}
	\includegraphics[width=8cm]{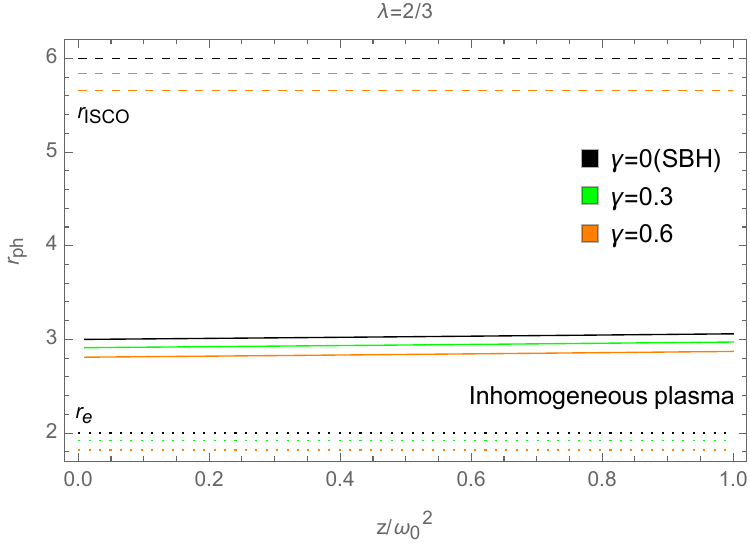}
	\caption{Variation of the photon sphere radius $r_{ph}$ of the KRBH with plasma frequency in homogeneous (left panel) and inhomogeneous (right panel) plasma for different LSB parameters. The dashed and dotted lines represent the ISCO radius $r_{ISCO}$ and event horizon radius $r_e$ of the KRBH, respectively. In the top panel, the black, red, and blue lines correspond to $\lambda=0, 1/2$ and 2/3 for fixed $\gamma=0.3$. In the bottom panel, the black, green, and orange lines correspond to $\gamma=0, 0.3$ and 0.6 for fixed $\lambda=2/3$.}
	\label{fig5110}
\end{figure}

In this section, we consider the KRBH surrounded by a cold, dust-like (pressure $P=0$) and non-magnetized plasma medium. The Hamiltonian for photons moving around the KRBH can be written as \cite{2207.06994}:
\begin{equation}
	\mathcal{H}= \frac{1}{2} \left[ g^{\mu\nu} p_{\mu} p_{\nu} + {\omega_p(r)}^2 \right],
	\label{eq536}
\end{equation}
where $\omega_p(r) = \frac{4\pi e^2}{m_e} N_e(r)$ is the plasma frequency, $ N_e$ is the particle number density, $e$ and $m_e$ represent the electron charge and mass, respectively. If the background spacetime is asymptotically flat, the photon frequency observed by a stationary observer at infinity is \cite{1507.04217v2}:
\begin{equation}
	\omega(r) = \frac{\omega_0}{\sqrt{A(r)}},
	\label{425}
\end{equation}
where $\omega_0$ is the photon energy at spatial infinity. Moreover, the photon frequency must be much greater than the plasma frequency ($\omega \gg \omega_p(r)$) for photons to propagate in the plasma medium \cite{1507.04217v2}. The $\omega_p(r)^2$ is constant in the case of the homogeneous plasma, while for the inhomogeneous plasma, it is represented as: 
\begin{equation}
	\omega_p(r) ^2= \frac{z}{r},\label{eq538}
\end{equation}
where $z$ is the free constant parameter. Using $\mathcal{H}= 0$, we derive the equation of motion for photons around the KRBH in the plasma medium:
\begin{equation}
	\frac{dr}{d\phi} =r \sqrt{1-\frac{2M}{r}+\frac{\gamma}{r^{2 / \lambda}}} \sqrt{r^2 ( \frac{1}{1-\frac{2M}{r}+\frac{\gamma}{r^{2 / \lambda}}} - \frac{\omega_p^2}{\omega_0^2} ) \frac{\omega_0^2}{p_\phi^2} - 1}.
	\label{eq537}
\end{equation}
Due to the presence of plasma, we redefine the impact parameter $b=\frac{p_\phi}{\omega_0}$. Defining $h(r)^2=r^2( \frac{1}{1-\frac{2M}{r} + \frac{\gamma}{r^{2 / \lambda}}} - \frac{\omega_p^2}{\omega_0^2})$, the radius of the photon sphere can be determined by the condition $\left. \frac{d(h(r)^ 2)}{dr} \right|_{r=r_{ph}} = 0$ \cite{hr}. Therefore, the photon sphere radius $r_{ph}$ of the KRBH in the plasma satisfies the expression:
\begin{equation}
	\frac{2 r_{ph}^{\frac{2+\lambda}{\lambda}} \left(-3 M r_{ph}^{2/\lambda} \lambda + r_{ph}^{\frac{2+\lambda}{\lambda}} \lambda + r_{ph} \gamma (1 + \lambda)\right)}{\left(-2 M r_{ph}^{2/\lambda} + r_{ph}^{\frac{2+\lambda}{\lambda}} + r_{ph} \gamma\right)^2 \lambda} - \frac{2 \omega_p^2}{\omega_{0}^2}=0.
	\label{425}
\end{equation}
According to Eq. (\ref{425}), Fig. \ref{fig5110} displays the variation of the photon sphere radius $r_{ph}$ of KRBH in both homogeneous and inhomogeneous plasma as a function of plasma frequency, for different values of the LSB parameters $\gamma$ and $\lambda$. From Fig. \ref{fig5110}, we observe that in the homogeneous plasma (left panel), compared to the case with no plasma ($\omega_p^2/\omega_0^2=0$), the photon sphere radius $r_{ph}$ of KRBH increases significantly as the plasma frequency $\omega_p^2/\omega_0^2$ increases. In the inhomogeneous plasma (right panel), the increase in $z/\omega_0^2$ has a relatively small effect on the photon sphere radius of KRBH. Moreover, regardless of whether the plasma distribution is homogeneous and inhomogeneous, the trend of the photon sphere radius $r_{ph}$ remains generally similar. However, as the LSB parameters $\lambda$ and $\gamma$ increase, the difference in the photon sphere radius between KRBH and SBH (black curve) gradually increases, with the photon sphere radius of KRBH always being smaller than the corresponding value of SBH.

\subsection{Optical Appearance and Shadow of KRBH in Plasma}
\begin{figure}[H]
	\centering
	\includegraphics[width=8cm]{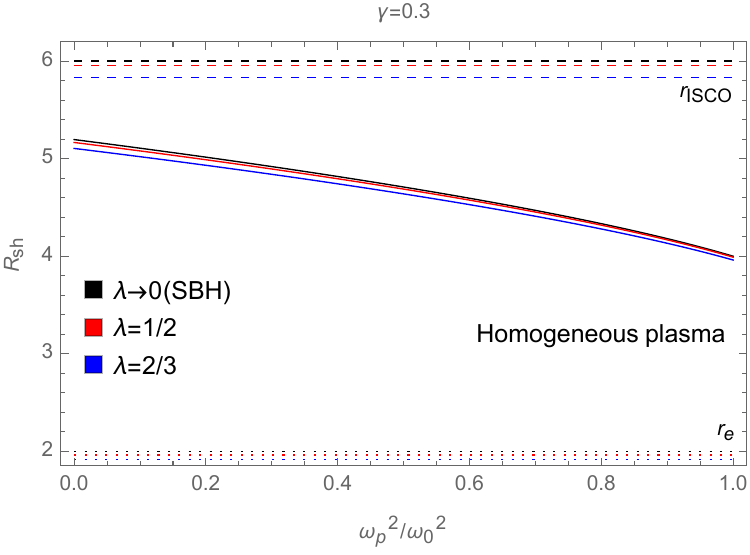}
	\includegraphics[width=8cm]{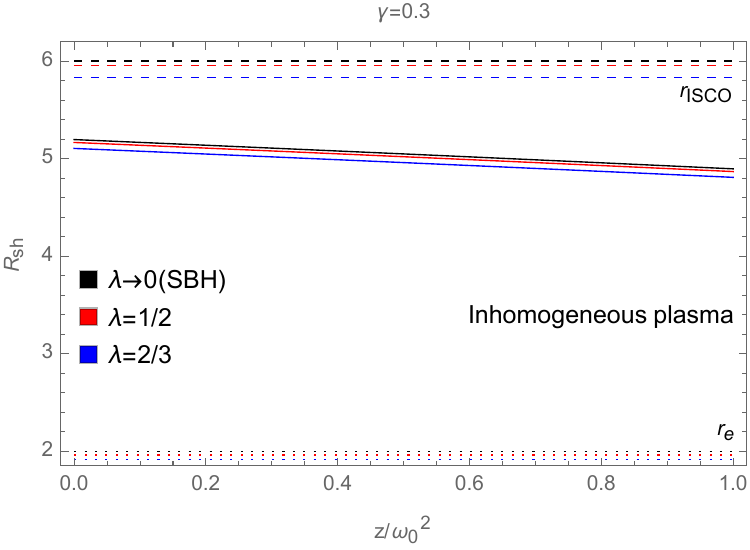}
	\includegraphics[width=8cm]{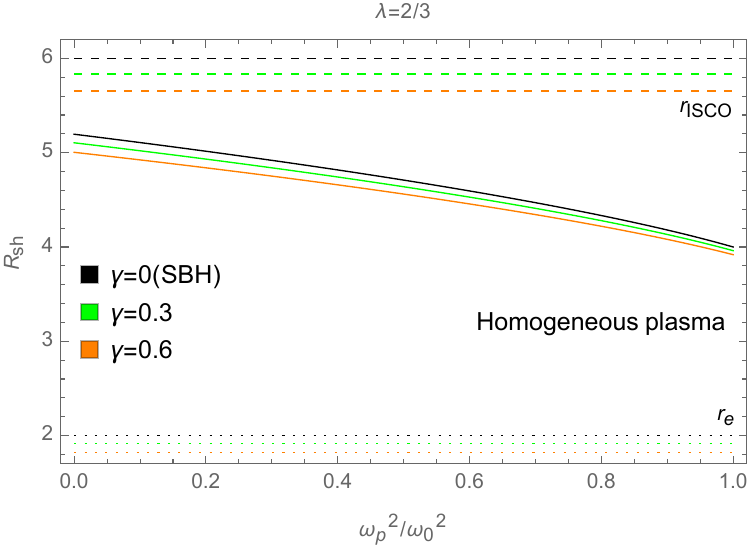}
	\includegraphics[width=8cm]{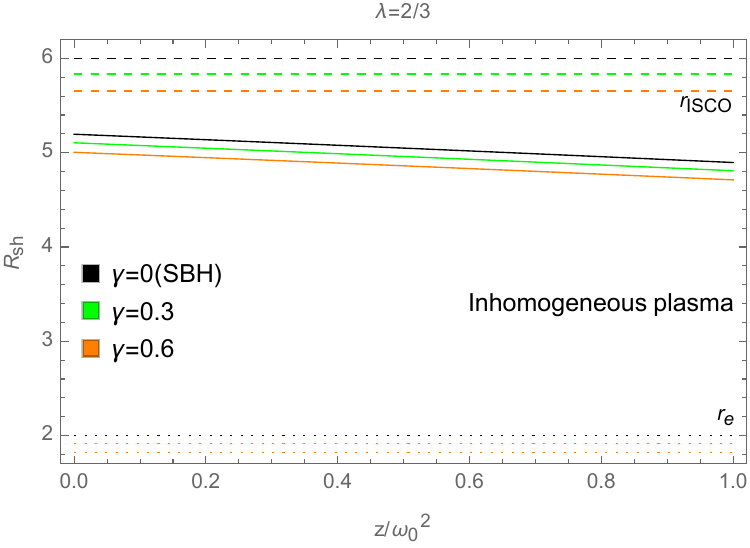}
	\caption{Variation of the KRBH shadow radius $R_{sh}$ as a function of plasma frequency in homogeneous (left panel) and inhomogeneous (right panel) plasma for different LSB parameters, as shown by solid lines. The dashed and dotted lines represent the ISCO radius $r_{ISCO}$ and event horizon radius $r_e$ of the KRBH, respectively. In the top panel, the black, red, and blue lines correspond to $\lambda=0, 1/2$ and 2/3 for fixed $\gamma=0.3$. In the bottom panel, the black, green, and orange lines correspond to $\gamma=0, 0.3$ and 0.6 for fixed $\lambda=2/3$.}
	\label{fig512}
\end{figure}

When there is plasma surrounding the KRBH, for a distant observer ($r_{obs} \to \infty$), the BH shadow radius is rewritten as \cite{2304.03660}:
\begin{equation}
	R_{sh} = \sqrt{r_{ph}^2 \left( \frac{1}{A(r_{ph})} - \frac{\omega_p^2}{\omega_0^2} \right) A(r_{obs})}.
	\label{eq539}
\end{equation}
Based on Eq. (\ref{eq539}), Fig. \ref{fig512} shows the variation of the KRBH shadow radius $R_{sh}$ with the LSB parameters $\lambda$ and $\gamma$ in both homogeneous (left panel) and inhomogeneous (right panel) plasma. From the figure, we can observe that as the plasma frequency $\omega_p^2/\omega_0^2$ or $z/\omega_0^2$ increases, the shadow of the KRBH gradually shrinks inward. Moreover, as the LSB parameters $\gamma$ or $\lambda$ increases, the shadow radius of the KRBH $R_{sh}$ decreases and is always smaller than the corresponding value for the SBH (black curve). Notably, for the same LSB parameter, the KRBH has a larger shadow radius in the inhomogeneous plasma, compared to the homogeneously plasma. This suggests that the KRBH will capture more photons in the inhomogeneous plasma. 

\begin{figure}[H]
	\centering
	\includegraphics[width=8cm]{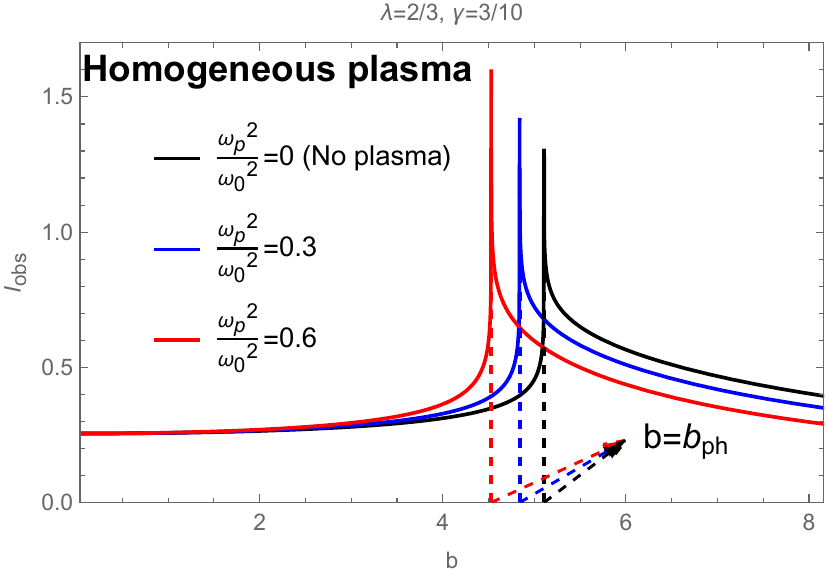}
	\includegraphics[width=8cm]{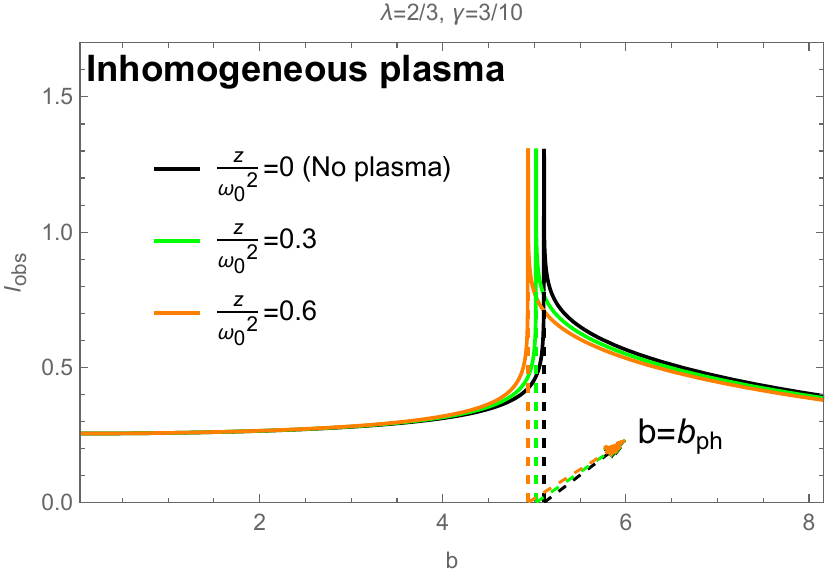}
	\caption{Variation of the total observed intensity of KRBH with impact parameter in the presence of different plasma frequencies, for $\lambda$=$\frac{2}{3}$ and $\gamma=0.3$.}
	\label{fig510}
\end{figure}

\begin{figure}[H]
	\centering
	\includegraphics[width=5cm]{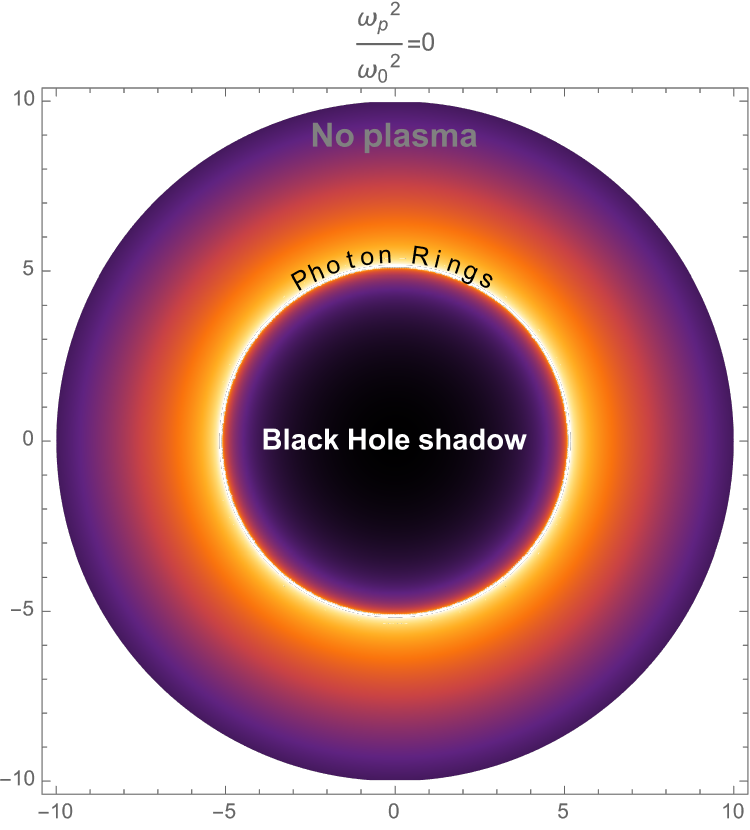}
	\includegraphics[width=5cm]{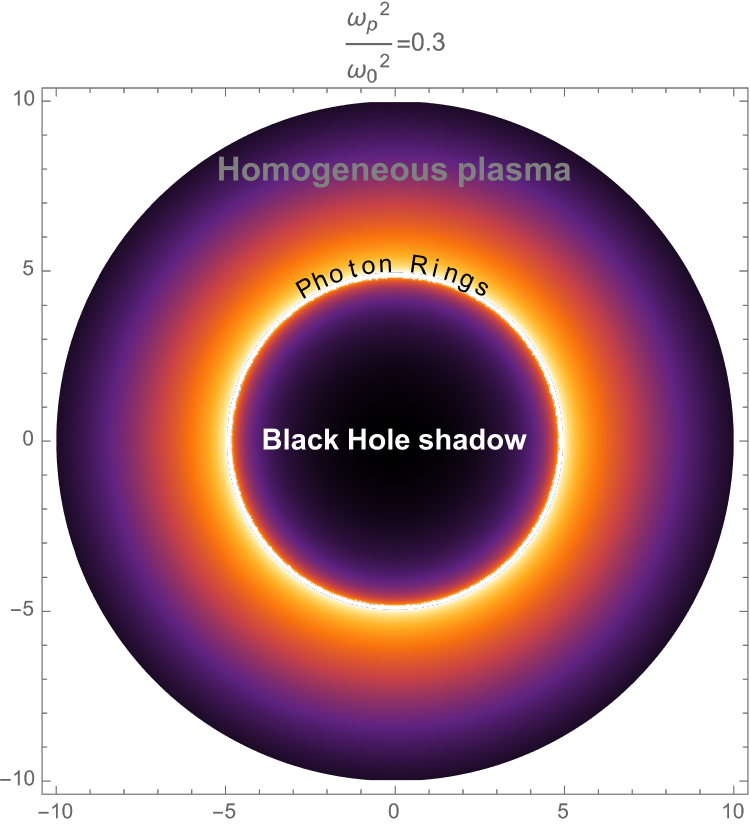}
	\includegraphics[width=5cm]{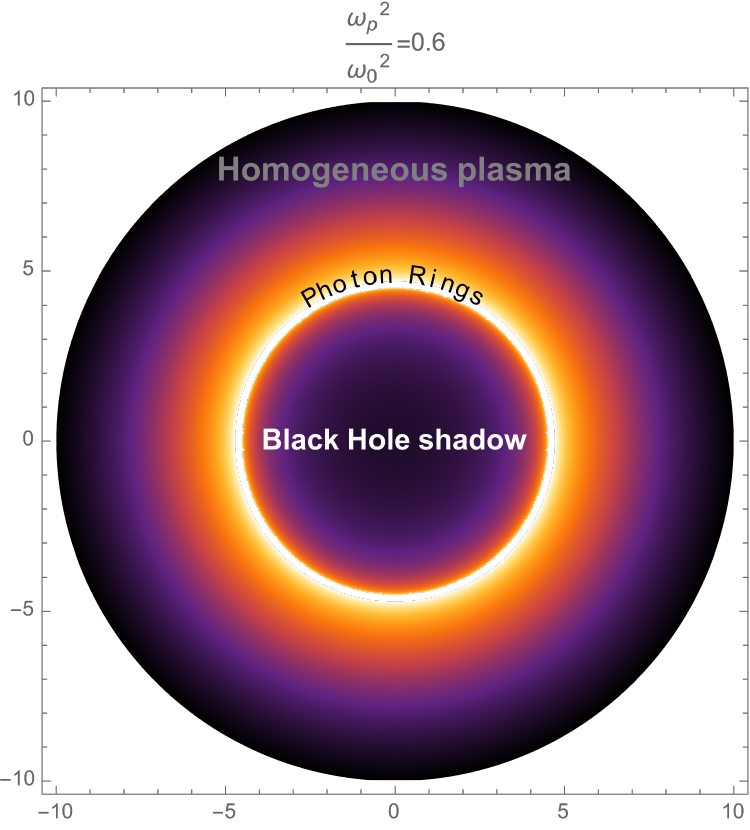}
	\includegraphics[width=5cm]{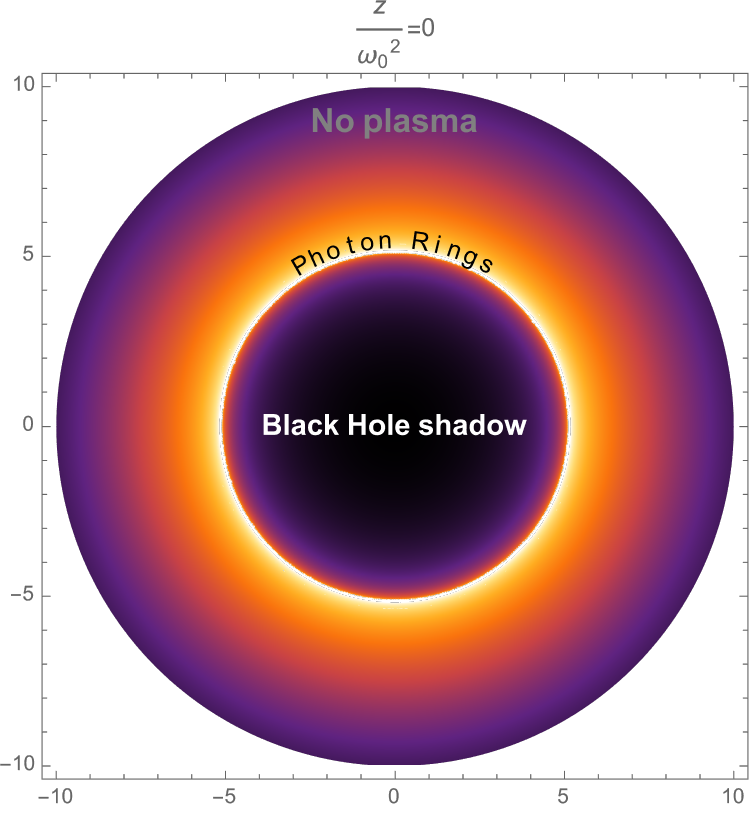}
	\includegraphics[width=5cm]{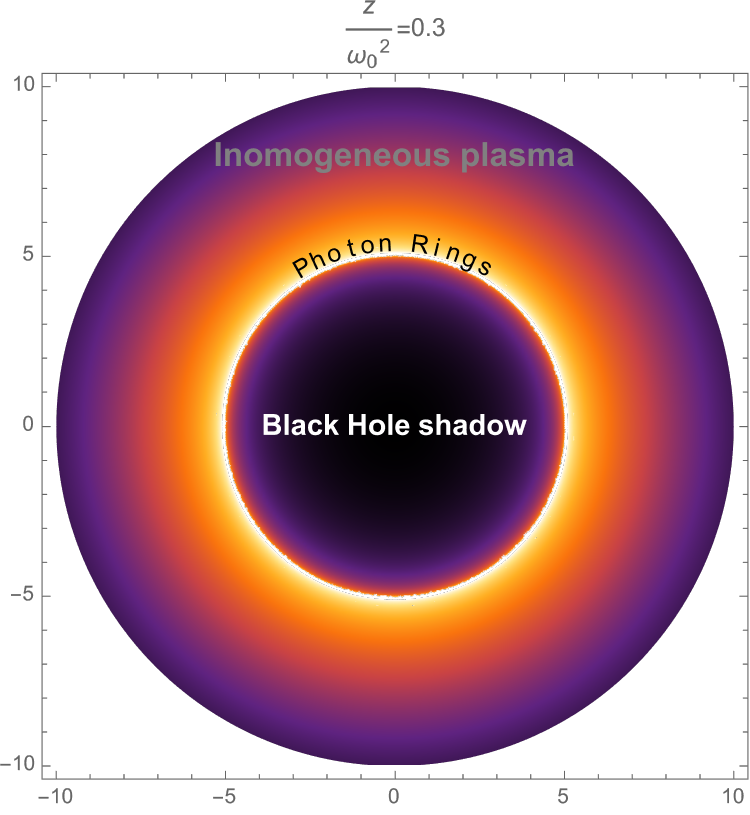}
	\includegraphics[width=5cm]{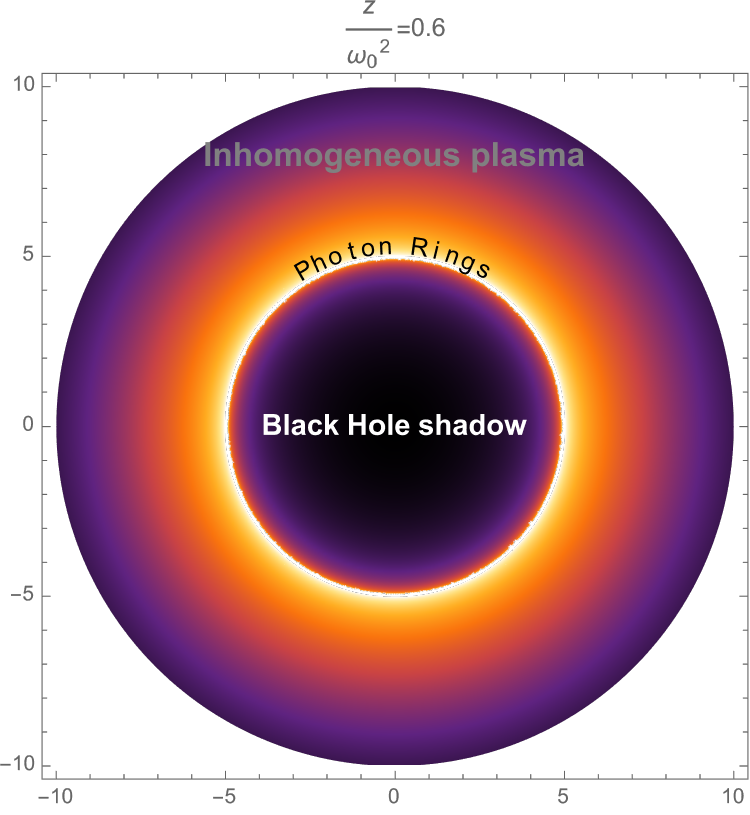}
	\caption{Optical appearance of KRBH in the static spherical accretion model, in homogeneous (top panel) and inhomogeneous (bottom panel) plasma, for $\lambda$=$\frac{2}{3}$ and $\gamma=0.3$.}
	\label{fig514}
\end{figure}

Next, we explore the observational features of the KRBH in the static spherical accretion model under both homogeneous and inhomogeneous plasma, respectively. The specific intensity observed by a distant observer can be given by the Eqs. (\ref{eq432}-\ref{eq433}) and (\ref{eq537}) \cite{1304.5691}:
\begin{equation}
	I_{obs} = \int \frac{A(r)^{3/2}}{r^2} \sqrt{\frac{1}{A(r)} + \frac{b^2}{(h(r)^2 - b^2) A(r)}} dr.
	\label{eq322}
\end{equation}
In Fig. \ref{fig510}, we show the effect of different plasma frequencies on the observed intensity of the KRBH for the LSB parameter $\lambda=\frac{2}{3}$, $\gamma=0.3$. When the plasma distribution is homogeneous (left panel), we find that the presence of plasma enhances the observed intensity peak of the KRBH, compared to the case with no plasma. This indicates that, in the static spherical accretion model, the shadow of the KRBH appears brighter due to the presence of plasma. Furthermore, the increase in the plasma frequency $\frac{\omega_p^2}{\omega_0^2}$ within the homogeneous plasma has a positive effect on the total observed intensity. However, in the inhomogeneous plasma (right panel), the effect of plasma on the KRBH observed intensity is relatively weak, with only minor changes compared to the left panel. Furthermore, in Fig. \ref{fig514}, we show the effect of the plasma on the optical appearance of KRBH under the spherical accretion model. From the figure, it can be observed that as the plasma frequency $\frac{\omega_p^2}{\omega_0^2}$ and $\frac{z}{\omega_0^2}$ increase, the brightness of the KRBH shadow gradually intensifies, with more noticeable effects in the homogeneous plasma. The presence of the plasma makes the optical appearance of the KRBH more pronounced, thus facilitating the distinction between KRBH and SBH.

\section{Shadows and Photon Rings of KRBH Illuminated by Thin Accretion Disk}
In the universe, compact celestial objects usually form accretion disks by accreting surrounding matter. The accretion material can be seen as the light source that illuminates it. For simplicity, in this section, we consider the simple accretion disk model, where the material surrounding the BH forms a geometrically and optically thin disk structure located on the equatorial plane. Furthermore, we assume that the radiation emitted by the accretion disk is isotropic, and a distant static observer is located along the northern polar direction.

\subsection{Three Types of Rays from the Accretion Disk Radiation Around the KRBH}
Rays emitted by the accretion disk intersect with the disk several times before reaching the observer screen. One can introduce the total number of photon orbits $n=\frac{\phi}{2 \pi}$ to classify photons orbiting around the BH \cite{2211.04263}. Using the orbital equations (\ref{eq211}-\ref{eq212}) and considering the transformation $u=\frac{1}{r}$, the total change in azimuthal angle for photons with $b<b_{ph}$ is given by:
\begin{equation}
	\phi=\int_{0}^{u_{e}} \frac{1}{\left[\frac{1}{b^{2}}-u^{2} A(u)\right]^{1 / 2}} d u,
	\label{eq222}
\end{equation}
where $u_{e}=\frac{1}{r_{e}}$. For rays with $b>b_{ph}$, the total change in azimuthal angle is:
\begin{equation}
	\phi=2 \int_{0}^{u_{t}} \frac{1}{\left[\frac{1}{b^{2}}-u^{2} A(u)\right]^{1 / 2}} d u,
	\label{eq223}
\end{equation}
where $u_{t}=\frac{1}{r_{t}}$ and $r_{t}$ is the minimum radial distance of the photon trajectory from the BH. The total number of orbits $n(b)$ satisfies the condition \cite{1906.00873, 2008.00657}:
\begin{equation}
	n(b) = \frac{2m - 1}{4}, \quad m = 1, 2, 3, \ldots
	\label{eq224}
\end{equation}
where $m$ is the number of times the ray from infinity passes through the equatorial plane of the KRBH. For each given value of $m$, Eq. (\ref{eq224}) will have two solutions $b_{m}^{-}$ and $b_{m}^{+}$. The rays can be classified into the following three types \cite{2211.04263}:

(1) Direct : $0 < n(b) < 3/4$, $b \in \left(0, b_{2}^{-}\right) \cup \left(b_{2}^{+}, \infty\right)$,

(2) Lensed ring : $3/4 < n(b) < 5/4$, $b \in \left(b_{2}^{-}, b_{3}^{-}\right) \cup \left(b_{3}^{+}, b_{2}^{+}\right)$,

(3) Photon ring : $n(b) > 5/4$, $b \in \left(b_{3}^{-}, b_{3}^{+}\right)$.

In Table \ref{tab3} and Table \ref{tab4}, we show the boundary values of the impact parameter for different values of the parameters $\lambda$ and $\gamma$. From the calculations, we observe that with the increase in both LSB parameters, the regions of direct emission, lensed ring, and photon ring all gradually expand. Additionally, the region of direct emission around the KRBH is larger than the regions of the lensed ring and photon ring.

\begin{table}[H]
	\centering
	\caption{Boundary values of the impact parameter for the KRBH at different values of $\lambda$, when $\gamma=0.3$.}
	\begin{tabular}{|c|c|c|c|c|c|}
		\hline
		&$b_1^-$  &$b_2^-$  &$b_3^-$  &$b_3^+$  &$b_2^+$ \\
		\hline
		$\lambda=0$(SBH)&2.848  &5.015  &5.188  &5.228  &6.168 \\
		\hline      
		$\lambda=\frac{1}{2}$&2.798  &4.974  &5.158  &5.201  &6.156 \\
		\hline       
		$\lambda=\frac{2}{3}$&2.743  &4.906  &5.099  &5.046  &6.057 \\
		\hline            
	\end{tabular}
	\label{tab3}
\end{table}

\begin{table}[H]
	\centering
	\caption{Boundary values of the impact parameter for the KRBH at different values of $\gamma$, when $\lambda=\frac{2}{3}$.}
	\begin{tabular}{|c|c|c|c|c|c|}
		\hline
		&$b_1^-$  &$b_2^-$  &$b_3^-$  &$b_3^+$  &$b_2^+$ \\
		\hline
		$\gamma=0$(SBH) &2.848  &5.015  &5.188  &5.228  &6.168 \\
		\hline
		$\gamma=0.3$&2.743  &4.906  &5.095  &5.141  &6.114 \\
		\hline        
		$\gamma=0.6$&2.614  &4.776  &4.991  &5.046  &6.057 \\
		\hline                  
	\end{tabular}
	\label{tab4}
\end{table} 

\subsection{Optical Observational Features of the KRBH with Thin Accretion Disk Model}
Next, we explore the optical appearance of the KRBH under three thin accretion disk models. The specific intensity $I_{obs}$ received by a distant observer at the north pole is given by \cite{2307.1236v2}:
\begin{equation}
	I_{obs}(r) = g^{3} I_{em}(r),
	\label{eq225}
\end{equation}
where $I_{em}$ is the specific intensity of the emitted rays. The specific intensity of photons around the KRBH is expressed as:\begin{equation}
	I_{obs}(r) = A(r)^{3/2} I_{em}(r) = \left(1 - \frac{2M}{r} + \frac{\gamma}{r^{2/\lambda}}\right)^{3/2} I_{em}(r).
	\label{eq226}
\end{equation}
By integrating over all observed frequency ranges, the total observed specific intensity of the KRBH is given by:
\begin{equation}
	I_{obs}^{total}(r) = \int I_{obs}(r) d\nu_{obs} = \left(1 - \frac{2M}{r} + \frac{\gamma}{r^{2/\lambda}}\right)^2 I_{em}^{total}(r),\label{eq227}
\end{equation}
where $I_{em}^{total}(r) = \int I_{em}(r) d\nu_{em}$ is the total emitted specific intensity from the accretion disk, and $\nu_{em}$ and $\nu_{obs}$ represent the emit photon frequency and observed photon frequency, respectively. Each intersection of the photon with the accretion disk adds an additional brightness. Therefore, the total observed intensity is \cite{2304.10015}:
\begin{equation}
	I_{obs}^{total}(b) = \sum \left(1 - \frac{2M}{r} + \frac{\gamma}{r^{2/\lambda}}\right)^2 I_{em}^{total}(r) \Big|_{r=r_m(b)},
	\label{eq228}
\end{equation}
where $r_m(b)$ is the transfer function, and its specific expression is:
\begin{equation}
	r_m(b) = \frac{1}{u\left(b, \frac{(2m-1)\pi}{2}\right)}, \quad b \in \left(b_m^-, b_m^+\right),
	\label{540}
\end{equation}
which reveals the relationship between the photon impact parameter $b$ and the radial coordinate of the $m$th intersection with the accretion disk \cite{2206.12820}.

As previously mentioned, we adopt a geometrically thin and optically thin accretion disk model located in the equatorial plane surrounding the KRBH. Although the models considered in this work are relatively simple, they are still comparable to more realistic accretion disk models, such as the Novikov–Thorne (NT) model. The NT model considers relativistic effects such as gravitational redshift, Doppler boosting, light bending, with assuming the existence of viscous magnetic and turbulent stresses, etc. Although the NT model has some slightly different versions, its typical assumptions \cite{NTISCO}, for example, non-self-gravitating (neglecting the influence of the disk mass on the background metric), geometrically thin, and neglecting fluid inhomogeneities, magnetic fields, and radial heat transport, and assumes that radiation is emitted from the disk surface, are consistent with the main physical features of the models considered in this work. It should be noted that although some assumptions of the NT model can be relaxed, the condition of a geometrically thin disk is fundamental, and altering this assumption would impact the overall structure of the disk \cite{NTISCO}. For further details on the NT model, please refer to \cite{NTISCO,NT1,NT2}. Previous studies have shown that the observable features of BH images are primarily determined by the location and physical properties of the emitting material \cite{1906.00873}. Considering that the event horizon, photon sphere and ISCO are key geometrical radii that characterize the strong field structure of the BH, we selected the three toy models, as shown in following, to investigate the BH optical appearance, which have been extensively investigated in previous literature \cite{2307.1236v2,disk4,disk5,disk6}. Notably, the inner edge of the NT disk is located at the ISCO, which coincides with the initial emission region in our first toy model. 

The first model assumes that the emission intensity $I_{em}^{total}$ peaks at the ISCO, and the total emission specific intensity is a decay function of the power of second-order related to the radial coordinate. The mathematical expression is as follows:

\begin{equation}
	I_{em1}^{total}=\left\{\begin{array}{cc}
		\left[\frac{1}{r-\left(r_{ISCO}-1\right)}\right]^{2} & \quad r>r_{ISCO}, \\
		0 &  \quad r \leq r_{ISCO}.
	\end{array}\right. 
	\label{eq229}
\end{equation}

In the second model, the emitted rays originate at the photon sphere radius $r_{ph}$, where optically thin matter contributes to the formation of the photon ring structure \cite{disk1,disk2,disk3}, and the emission intensity decreases with increasing radial coordinates in the form of the following third power function:
\begin{equation}
	I_{e m 2}^{{total }}=\left\{\begin{array}{cc}
		{\left[\frac{1}{r-\left(r_{p h}-1\right)}\right]^{3}} &\quad r>r_{p h}, \\
		0 &\quad r \leq r_{p h}.
	\end{array}\right.
	\label{eq230}
\end{equation}

The third model emission profile begins at the event horizon $r_e$ and follows the form:
\begin{equation}
	I_{e m 3}^{{total }}=\left\{\begin{array}{cc}
		\frac{\frac{\pi}{2}-\arctan \left[r-\left(r_{I S C O}-1\right)\right]}{\frac{\pi}{2}-\arctan \left[r_{e}-\left(r_{I S C O}-1\right)\right]} &\quad r>r_{e}, \\
		0 &\quad r \leq r_{e}.
	\end{array}\right.
	\label{eq231}
\end{equation}
\begin{figure}[H]
	\centering
	\includegraphics[width=5cm]{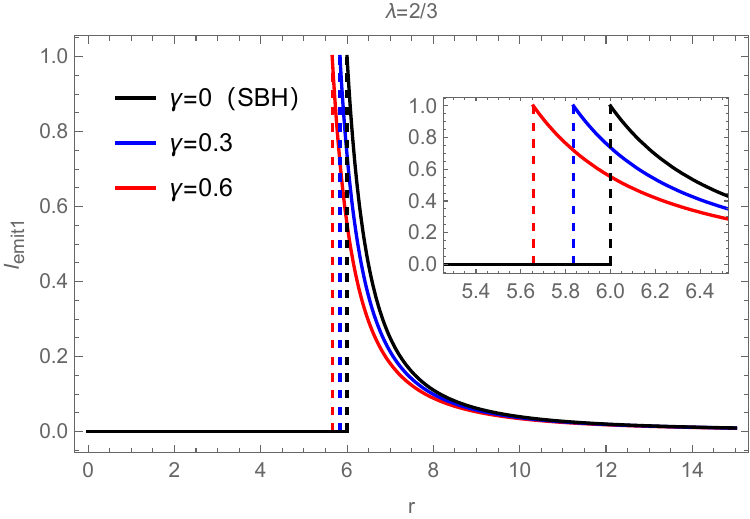}
	\includegraphics[width=5cm]{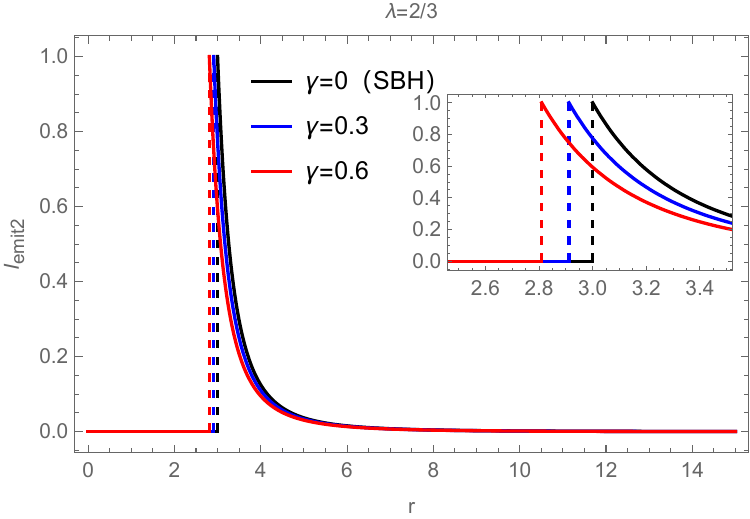}
	\includegraphics[width=5cm]{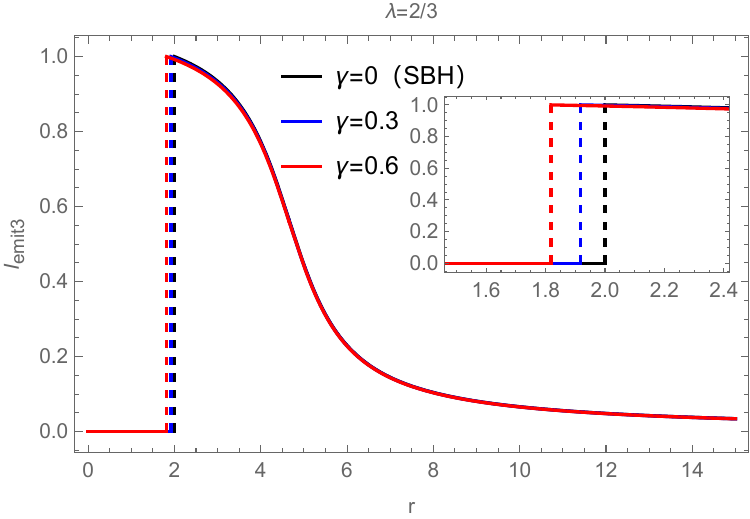}
	\includegraphics[width=5cm]{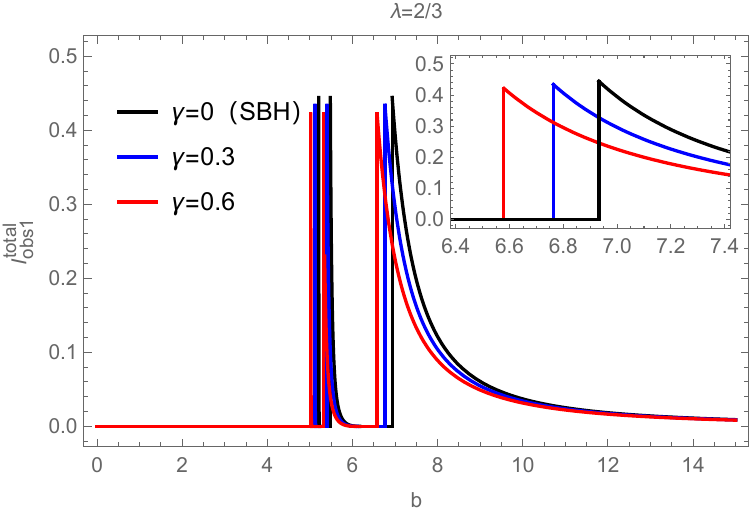}
	\includegraphics[width=5cm]{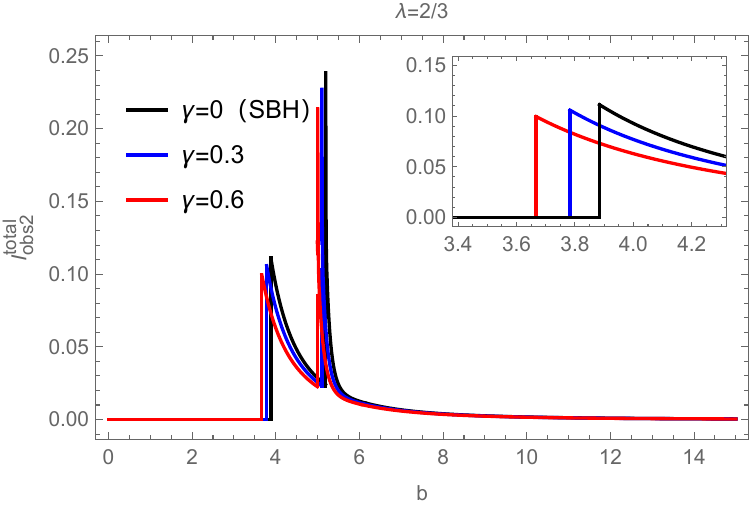}
	\includegraphics[width=5cm]{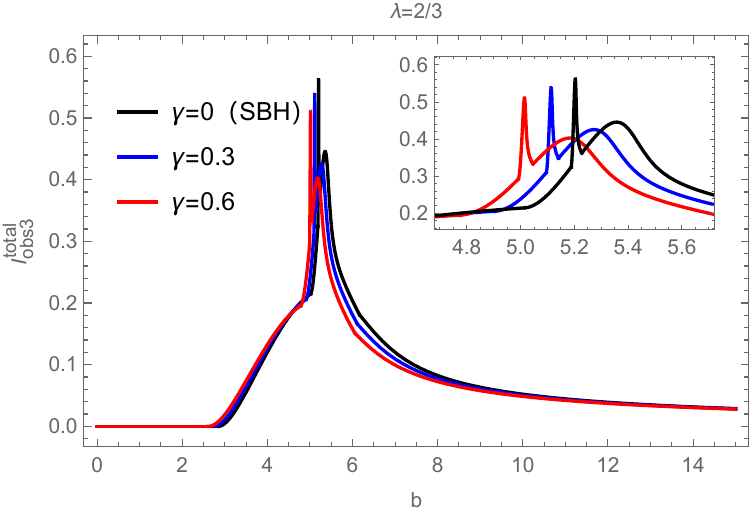}
	\caption{Top row: The total emitted intensity $I^{total}_{em}$ of the thin accretion disk as a function of the radial coordinate $r$. Bottom row: The dependence of the total observed intensity $I^{total}_{obs}$ on the impact parameter $b$. The black curve represents the SBH ($\gamma=0$), while the blue and red curves correspond to the KRBHs with $\gamma=0.3$ and $\gamma=0.6$, respectively. From left to right, the emission models correspond to $I_{em1}^{total}$, $I_{em2}^{total}$, and $I_{em3}^{ total}$, for $\lambda=2/3$.}
	\label{fig28}
\end{figure}

In Fig. \ref{fig28}, we show the curves of the emission intensity with radial coordinate $r$ and the trend of the total observed intensity with impact parameter $b$ for three accretion disk models. We choose the LSB parameter $\lambda=\frac{2}{3}$ and consider the three cases $\gamma=0$, 0.3 and 0.6, respectively. The first column of figures corresponds to the first accretion disk model, where the total observed intensity shows three peaks. When $\lambda=\frac{2}{3}$ and $\gamma=0.3$, the intensity of the direct emission starts to decrease at $b\approx6.762$, and the contribution of the lensed rings is mainly distributed in the range $5.401\lesssim b\lesssim6.113$. In addition, the photon ring produces an extremely narrow peak at $b\approx5.118$, which contributes less to the total observed intensity. The second column shows the case for model II. In this model, the emission regions of the photon ring and lensed ring overlap with the direct emission region, making them difficult to distinguish. Overall, the contributions of the lensed ring and photon ring to the total observed intensity are small, with the main contribution coming from the direct emission. The third column corresponds to the third accretion disk model, where the peaks observed intensity in the photon ring region. The lensed ring contribution is enhanced compared to the second model, but the main contribution still comes from direct emission, and the effect of the photon ring is negligible.
\begin{figure}[H]
	\centering
	\includegraphics[width=5cm]{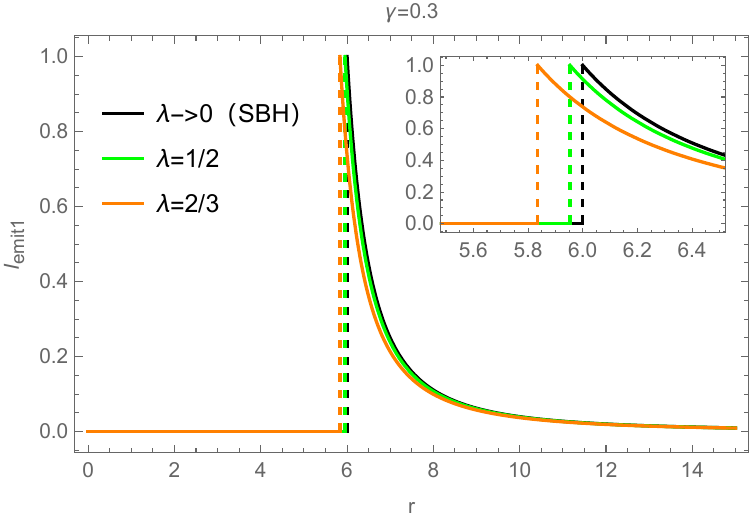}
	\includegraphics[width=5cm]{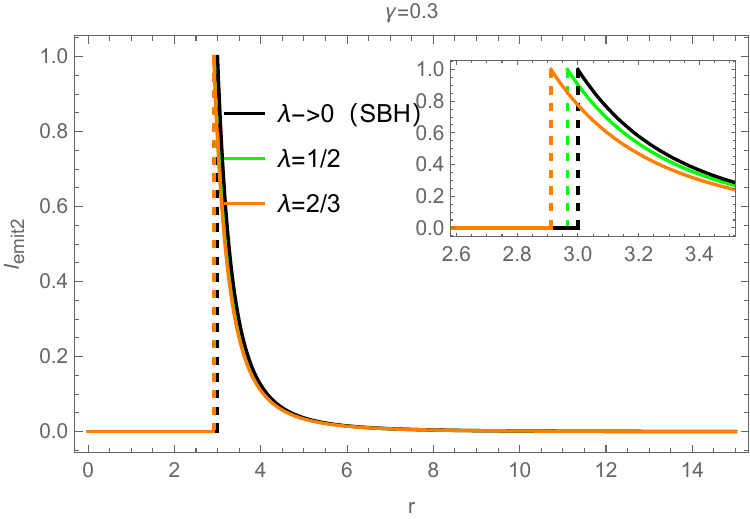}
	\includegraphics[width=5cm]{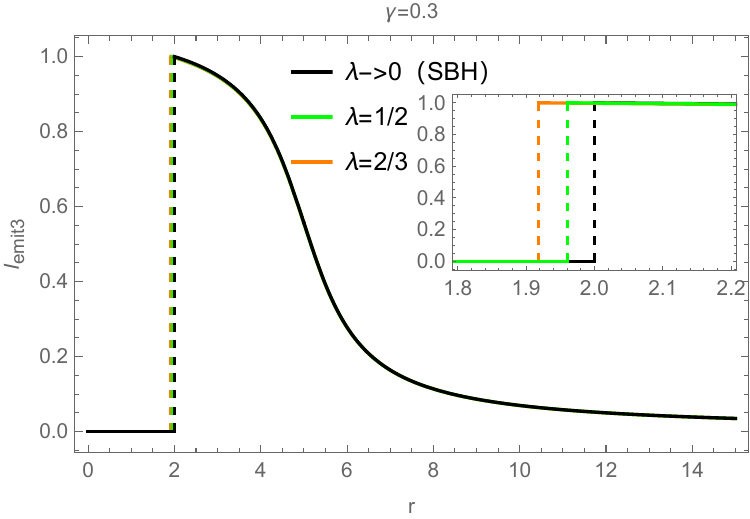}
	\includegraphics[width=5cm]{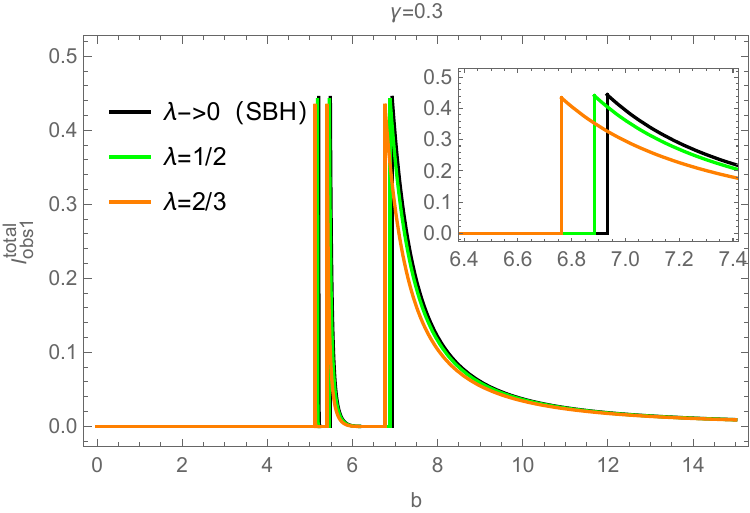}
	\includegraphics[width=5cm]{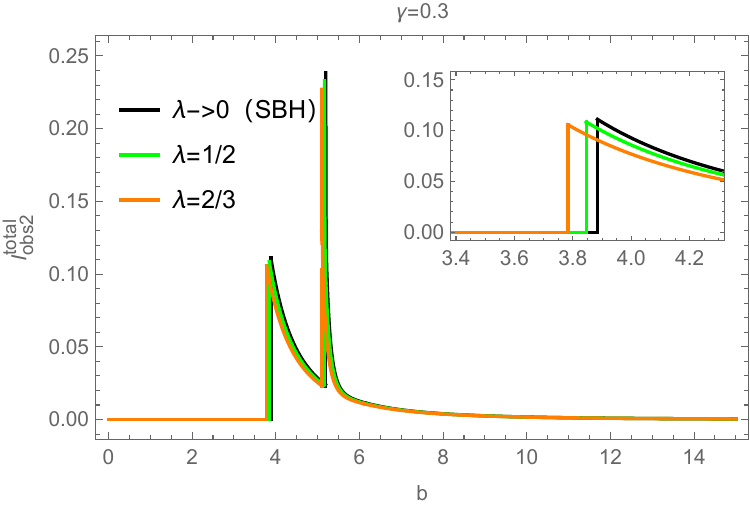}
	\includegraphics[width=5cm]{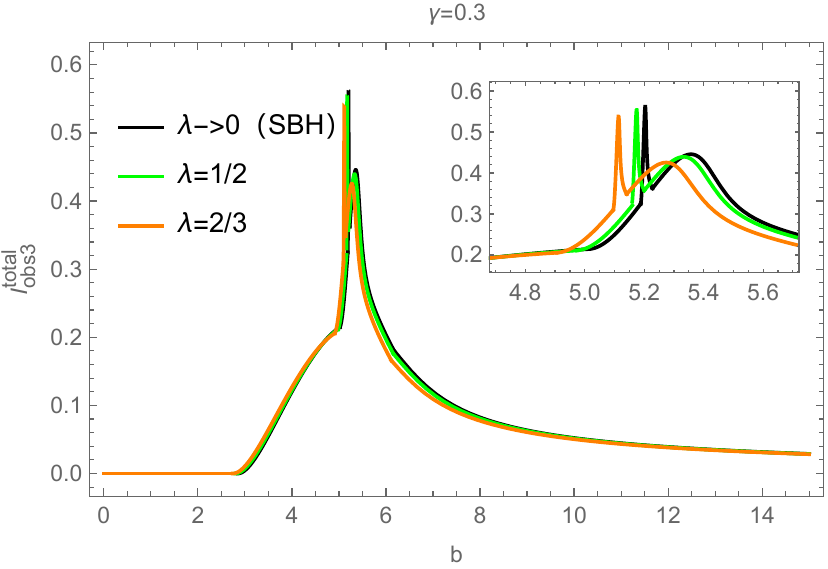}
	\caption{Top row: The total emitted intensity $I^{total}_{em}$ of the thin accretion disk as a function of the radial coordinate $r$. Bottom row: The dependence of the total observed intensity $I^{total}_{obs}$ on the impact parameter $b$. The black curve represents the SBH ($\lambda \to 0$), while the blue and red curves correspond to the KRBHs with $\lambda=\frac{1}{2}$ and $\lambda=\frac{2}{3}$, respectively. From left to right, the emission models correspond to $I_{em1}^{total}$, $I_{em2}^{total}$, and $I_{em3}^{ total}$, for $\gamma=0.3$.}
	\label{fig281}
\end{figure}
In Fig. \ref{fig281}, we select the LSB parameter $\gamma=0.3$ and consider three cases $\lambda \to 0$, $\frac{1}{2}$ and $\frac{2}{3}$, respectively. The figure shows the emission intensity curves for the three accretion disk models as a function of the radial coordinate $r$, along with the variation of the total observed intensity with respect to the impact parameter $b$. By analyzing the total observed intensity images of the KRBH under the three emission models, we observe that with the increase in the LSB parameters $\gamma$ and $\lambda$, the peaks observed intensity gradually decrease, and they are always lower than the corresponding SBH case. Moreover, as $\gamma$ and $\lambda$ increase, the thickness of the photon ring and lensed ring of the KRBH increases, but due to their high degree of demagnification, the contributions from the lensed ring and photon ring to the total observed intensity are negligible. Therefore, the total observed intensity of the KRBH mainly comes from direct emission. To visually distinguish the optical appearance of the KRBH in different accretion disk models, we present in Fig. \ref{fig282} the optical appearances of the SBH and the KRBH with LSB parameters $\lambda=\frac{2}{3}$ and $\gamma=0.3$ in three thin accretion disk models. The differences in the optical appearance allow for a clear distinction between the KRBH and SBH.

\begin{figure}[H]
	\centering
	\includegraphics[width=5cm]{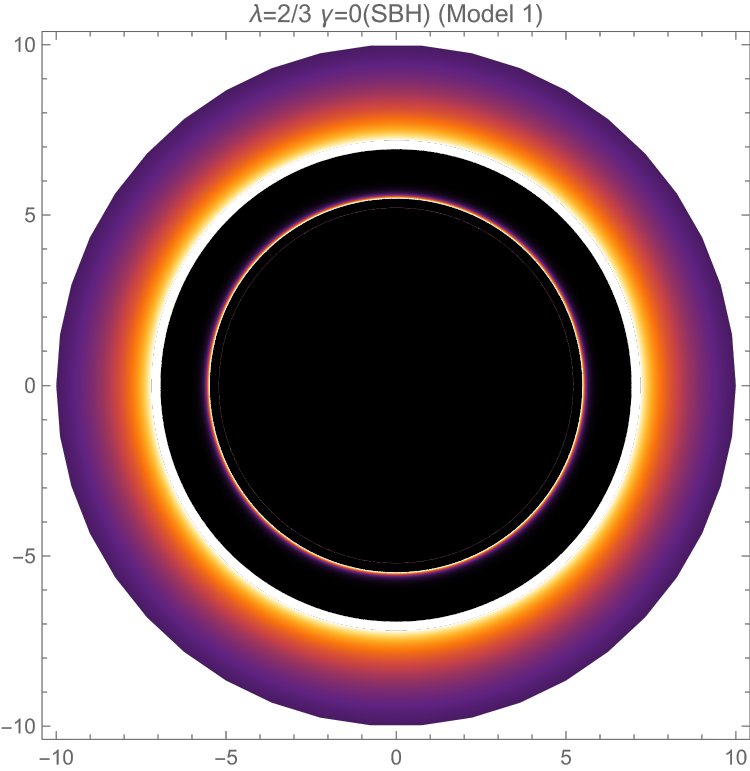}
	\includegraphics[width=5cm]{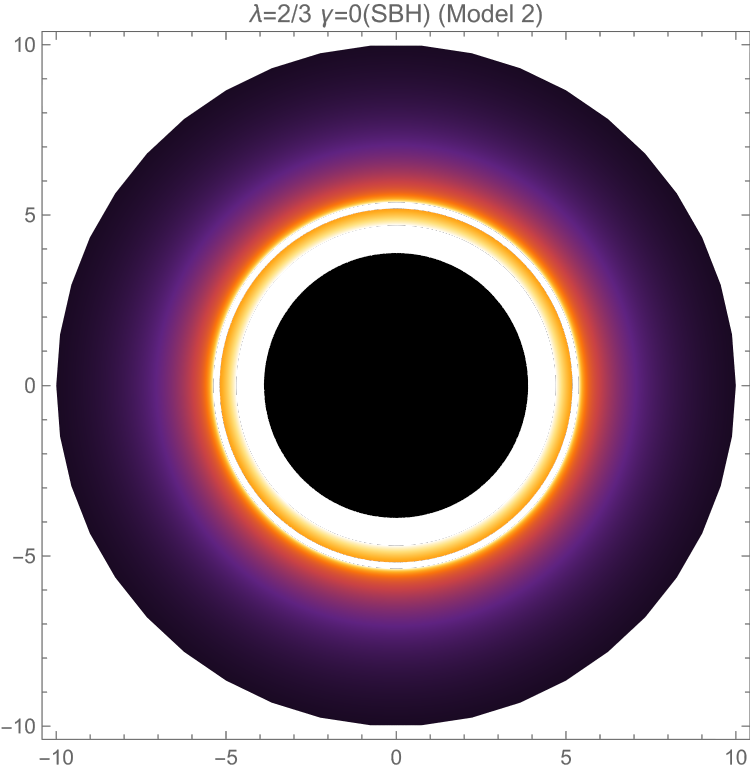}
	\includegraphics[width=5cm]{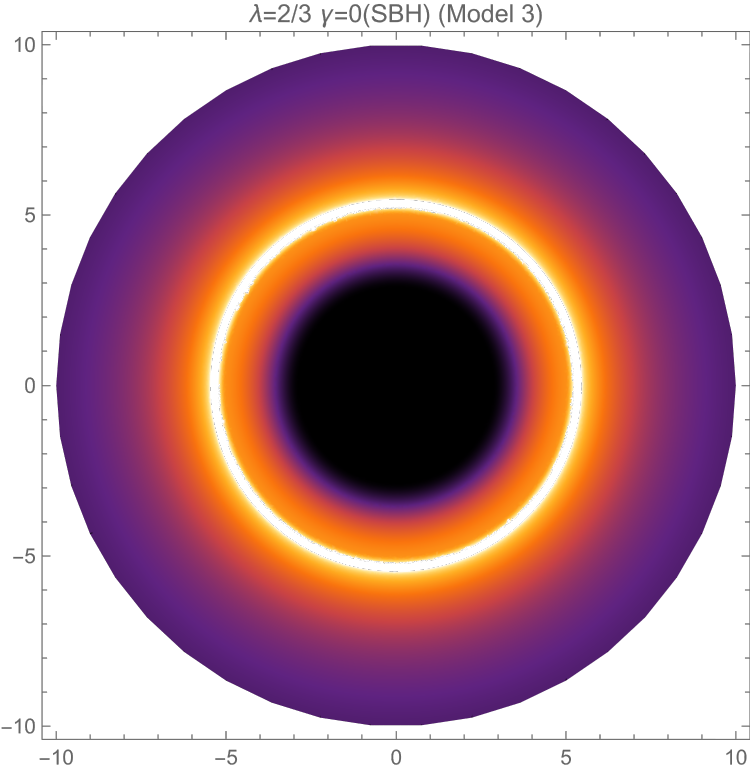}
	\includegraphics[width=5cm]{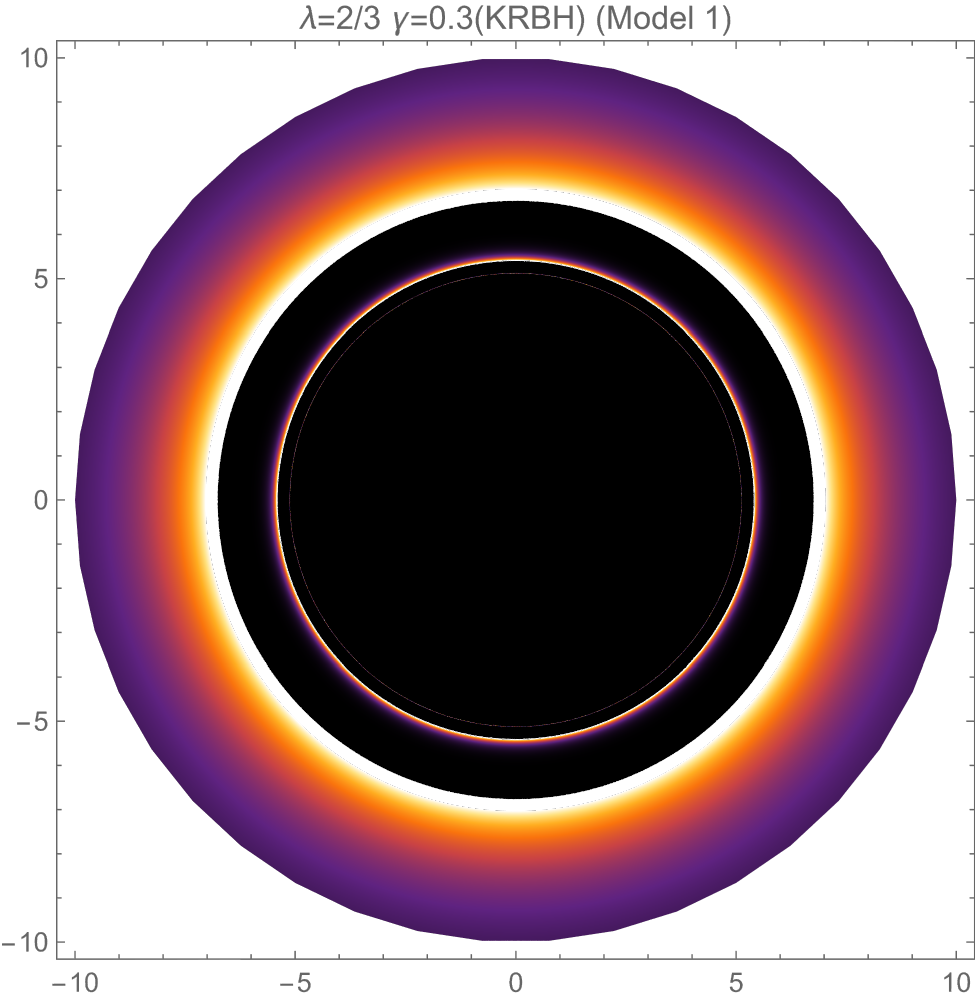}
	\includegraphics[width=5cm]{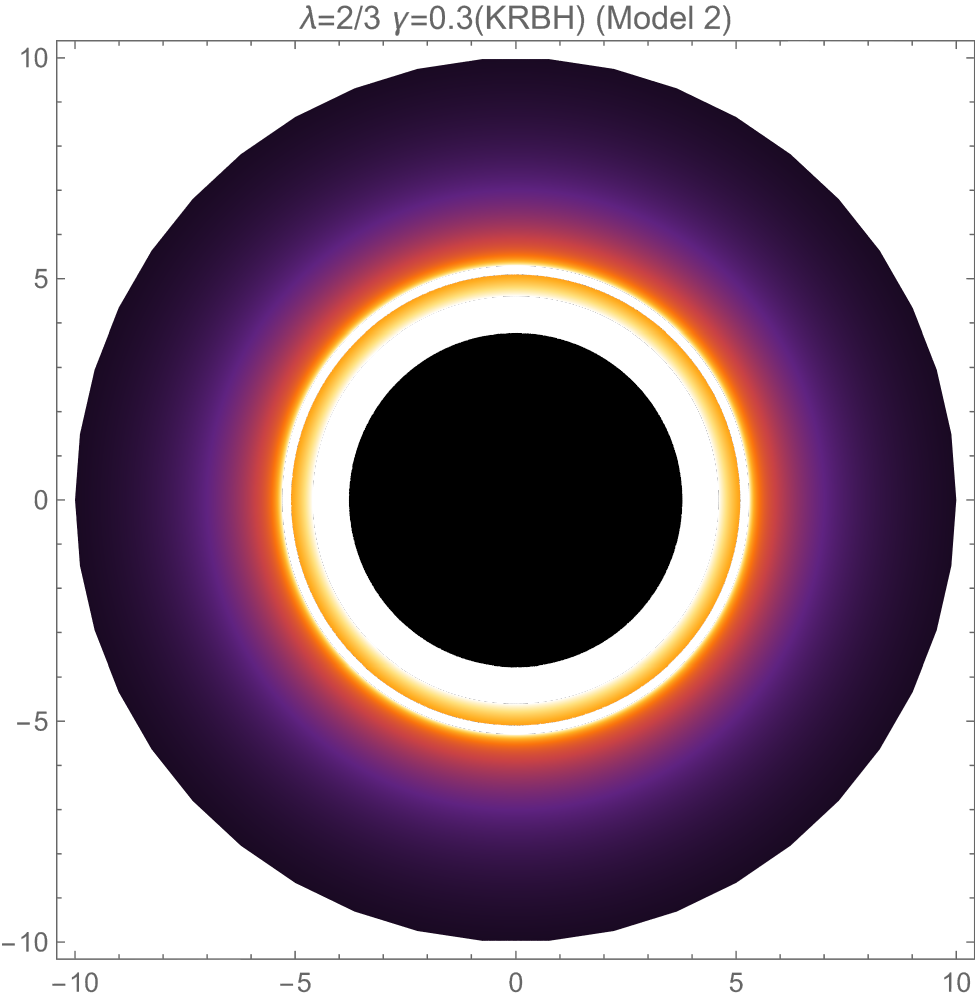}
	\includegraphics[width=5cm]{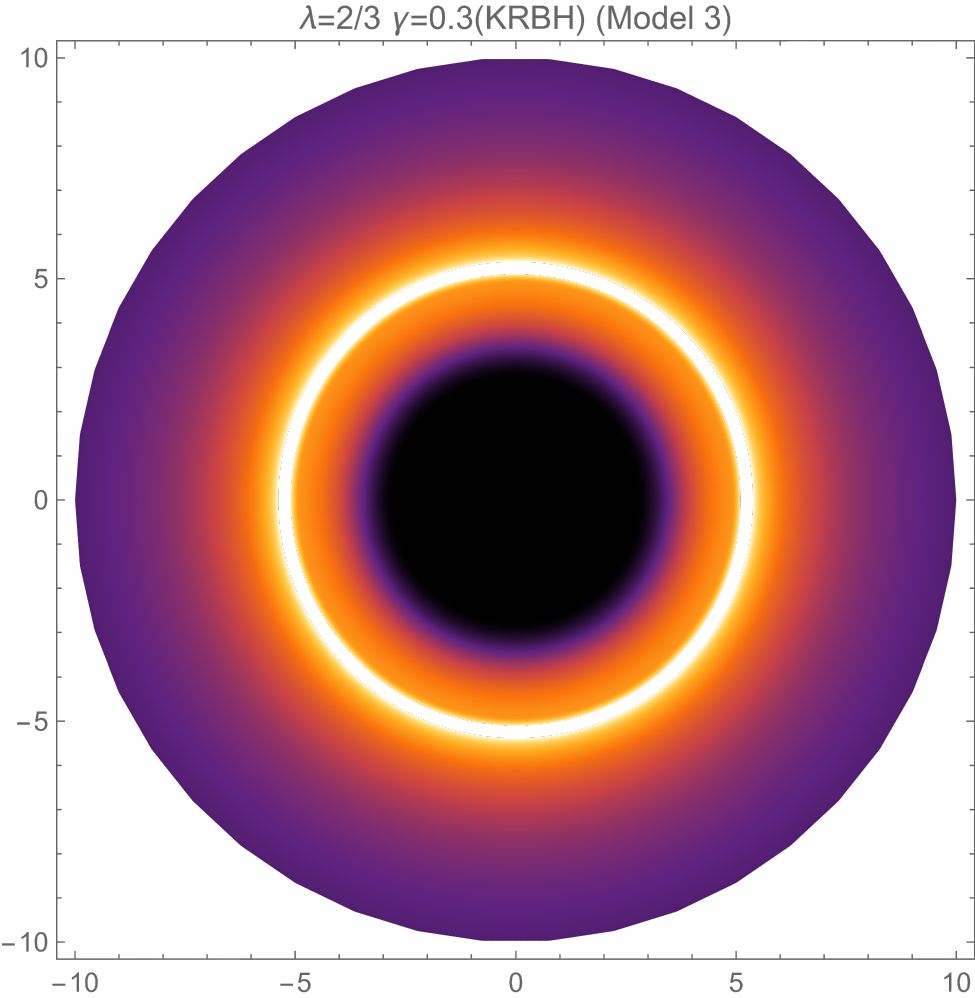}
	\caption{Optical appearances of the KRBH in three thin accretion disk models. The first row corresponds to the optical appearance of the SBH, while the second row shows the optical appearance of the KRBH with LSB parameters $\lambda=\frac{2}{3}$ and $\gamma=0.3$.}
	\label{fig282}
\end{figure}

\section{Conclusion}
In this paper, we investigated the effects of plasma presence and different accretion models on the shadow and optical appearance of static spherically symmetric BHs with the KR field. The KR field consists of a self-interacting antisymmetric 2-tensor. When the KR field non-minimally couples with the gravitational field and acquires a non-zero vacuum expectation value, Lorentz symmetry spontaneously breaks. We presented the static spherically symmetric BH solutions within the KR field, which include two LSB parameters, $\lambda$ and $\gamma$. By deriving the photon motion equations in the KRBH, we found that as $\lambda$ or $\gamma$ increase, the event horizon radius, photon sphere radius, and BH shadow radius of the KRBH decrease, always be smaller than the corresponding quantities for the SBH. Moreover, we constrained the LSB parameters using the shadow angular diameter data of the SgrA* released by the EHT. For $\lambda = 2/3$, the constraint on $\gamma$ was found to be $\gamma \lesssim 0.796$, and for $\gamma = 0.3$, the constraint on $\lambda$ was $\lambda \lesssim 0.981$.

The study revealed that when no plasma surrounds the KRBH, the total observed intensity of the KRBH in the static spherically symmetric accretion model increases with the impact parameter $b$, reaching a maximum at the critical impact parameter $b_{ph}$, and then decreases gradually. Notably, the increase in the LSB parameters $\lambda$ and $\gamma$ enhances the total observed intensity, meaning that the observed intensity of the KRBH is higher compared to that of the SBH. This characteristic can be used to distinguish between the two types of BHs. In addition, some of the radiating gas behind the KRBH is able to escape to infinity due to static spherically symmetric accretion flow, so that the observed intensity in the shadow region is not completely zero. 

Furthermore, the paper examined the shadow and optical appearance of the KRBH surrounded by plasma in the static spherically symmetric accretion model. We derived the photon motion equations in the presence of plasma around the KRBH. The results indicated that as the plasma frequency increases, the photon sphere radius of the KRBH increases, while the shadow radius decreases. In addition, under the condition of homogeneous plasma, the effect of varying the plasma frequency on the photon sphere radius and the BH shadow radius is more significant compared to the case of inhomogeneous plasma. For the optical appearance of the KRBH with a static spherically symmetric accretion model, we found that the presence of plasma, especially in the case of homogeneous plasma, enhances the peak of the observed intensity of the KRBH (compared to the inhomogeneous). The increase in the homogeneous plasma frequency positively affects the total observed intensity, indicating that the shadow of the KRBH becomes brighter due to the presence of the plasma.

Finally, we analyzed the optical appearance of the KRBH under three thin disk accretion models. It is shown that as the LSB parameter $\gamma$ increases, the peak of the observed intensity gradually decreases, remaining always smaller than the corresponding value for the SBH. Moreover, as $\gamma$ increases, the thickness of the photon ring and lensed ring around the KRBH increases. However, due to their high degree of demagnification, the contributions of the lensed ring and photon ring to the total observed intensity are extremely weak, meaning that the total observed intensity of the KRBH primarily comes from direct emission. Through the differences in optical appearance, we can effectively distinguish between the KRBH and the SBH.

\textbf{\ Acknowledgments }
The research work is supported by the National Natural Science Foundation of China (12175095,12075109 and 12205133), and supported by  LiaoNing Revitalization Talents Program (XLYC2007047).


\begin{thebibliography}{*}
\bibitem{116.061102}B.P. Abbott, R. Abbott, T.D. Abbott, et al. Observation of gravitational waves from a binary black hole merger[J]. Physical review letters, 2016, 116(6): 061102.
\bibitem{1906.11239}K. Akiyama, A. Alberdi,W. Alef, et al. First M87 event horizon telescope results. II. Array and instrumentation[J]. The Astrophysical Journal Letters, 2019, 875(1): L2.
\bibitem{1906.11240}T. Savolainen, Event Horizon Telescope Collaboration. First M87 Event Horizon Telescope Results. III. Data Processing and Calibration[J]. Astrophysical Journal Letters, 2019, 875(1): 3.
\bibitem{1906.11241}T. Savolainen, Event Horizon Telescope Collaboration. First M87 Event Horizon Telescope Results. IV. Imaging the Central Supermassive Black Hole[J]. Astrophysical Journal Letters, 2019, 875(1): 4.
\bibitem{2311.08679}I. Cho, G.Y. Zhao, T. Kawashima, et al. The Intrinsic Structure of Sagittarius A* at 1.3 cm and 7 mm[J]. The Astrophysical Journal, 2022, 926(2): 108.
\bibitem{Event}K. Akiyama, A. Alberdi, W. Alef, et al. First Sagittarius A* Event Horizon Telescope results. I. The shadow of the supermassive black hole in the center of the Milky Way[J]. The Astrophysical Journal Letters, 2022, 930(2): L12.

\bibitem{Synge} J.L. Synge. The escape of photons from gravitationally intense stars[J]. Monthly Notices of the Royal Astronomical Society, 1966, 131(3): 463-466.
\bibitem{Luminet} J.P. Luminet. Image of a spherical black hole with thin accretion disk[J]. Astronomy and Astrophysics, vol. 75, no. 1-2, May 1979, p. 228-235., 1979, 75: 228-235.
\bibitem{Bardeen} J.M. Bardeen. Les Houches Summer School of Theoretical Physics: Black Holes[J]. 1973.
\bibitem{review1}P.V.P. Cunha, N.A. Eiró, N.A. Herdeiro, et al. Lensing and shadow of a black hole surrounded by a heavy accretion disk[J]. Journal of Cosmology and Astroparticle Physics, 2020, 2020(03): 035.
\bibitem{review2} V.Perlick, O.Y. Tsupko. Calculating black hole shadows: Review of analytical studies. Phys. Rept., 947: 1–39[J]. arXiv preprint arXiv:2105.07101, 2022.
\bibitem{review3} S. Chen, J. Jing,  W.L. Qian, et al. Sci. China, Phys. Mech. Astron.(2022)[J]. arXiv preprint arXiv:2301.00113.
\bibitem{chin1}T.T. Sui, Q.M. Fu, W.D. Guo, The shadows of accelerating Kerr-Newman black hole and constraints from M87*. Physics Letters B 845, 10, 138135 (2023)
\bibitem{chin2}S. Hu, C. Deng, D. Li, X. Wu and E. Liang, Observational signatures of Schwarzschild-MOG black holes in scalar-tensor-vector gravity: shadows and rings with different accretions, Eur. Phys. J. C 82, 10, 885 (2022)
\bibitem{chin3} X. Wang, X. Kuang, Y. Meng, B. Wang, J. Wu, Rings and images of Horndeski hairy black hole illuminated by various thin accretions, Phys.Rev.D 107 (2023) 12, 124052.

\bibitem{PRD39} V.A. Kostelecký, S. Samuel. Spontaneous breaking of Lorentz symmetry in string theory[J]. Physical Review D, 1989, 39(2): 683. 
\bibitem{PRL63}V.A. Kostelecky, S. Samuel, Phys. Rev. Lett. 63, 224 (1989)
\bibitem{PRL66}V.A. Kostelecky, S. Samuel, Phys. Rev. Lett. 66, 1811 (1991)
\bibitem{0108061}J. Alfaro, H.A. Morales-Tecotl, L.F. Urrutia, Phys. Rev. Lett. 84,2318 (2000)
\bibitem{9809038}J. Alfaro, H.A. Morales-Tecotl, L.F. Urrutia, Phys. Rev. D 65,103509 (2002)
\bibitem{0411158}P. Horava, Phys. Rev. D 79, 084008 (2009)
\bibitem{PRL87}S.M. Carroll, J.A. Harvey, V.A. Kostelecky, C.D. Lane, T.Okamoto, Phys. Rev. Lett. 87, 141601 (2001)
\bibitem{0007031}T. Jacobson, D. Mattingly. Gravity with a dynamical preferred frame[J]. Physical Review D, 2001, 64(2): 024028.
\bibitem{9605088}V.A. Kostelecky,, and R. Potting. "Expectation Values, Lorentz Invariance, and CPT in the Open Bosonic String." Physics Letters B, arXiv preprint arXiv:hep-th/9605088v1 (1996).
\bibitem{0008252}V.A. Kostelecký, R. Potting. Analytical construction of a nonperturbative vacuum for the open bosonic string[J]. Physical Review D, 2001, 63(4): 046007.
\bibitem{1711.02273}R. Casana, A. Cavalcante, F.P. Poulis, et al. Exact Schwarzschild-like solution in a bumblebee gravity model[J]. Physical Review D, 2018, 97(10): 104001.
\bibitem{0312310v2}V.A. Kostelecký, M. Mewes. Lorentz and CPT violation in neutrinos[J]. Physical Review D, 2004, 69(1): 016005.
\bibitem{ORD396831989}V. A. Kostelecký, S. Samuel, Spontaneous breaking of Lorentz symmetry in string theory, Phys. Rev. D 39, 683 (1989).
\bibitem{Phys.Rev. D 401886 (1989}V.A. Kostelecký and S. Samuel, Gravitational phenomenology in higher dimensional theories and strings, Phys.Rev. D 40, 1886 (1989).
\bibitem{VAKSS}V.A. Kostelecký and S. Samuel, Phenomenological gravitational constraints on strings and higher dimensional.
\bibitem{Phys. Rev. D 74045001 (2006}Q.G. Bailey and V.A. Kostelecký, Signals for Lorentz violation in post-Newtonian gravity, Phys. Rev. D 74,045001 (2006).
\bibitem{theories} V.A. Kostelecký, S. Samueltheories, Phenomenological gravitational constraints on strings and higher-dimensional theories, Phys. Rev. Lett. 63, 224 (1989).
\bibitem{Phys.Rev.Lett.63224(1989}R. Bluhm, N.L. Gagne, R. Potting, and A. Vrublevskis,Constraints and stability in vector theories with spontaneous Lorentz violation, Phys. Rev. D 77, 125007 (2008).
\bibitem{1711.02273v1} R. Casana, A. Cavalcante, F.P. Poulis, E.B. Santos, Phys. Rev. D 97, 104001 (2018).
\bibitem{1811.08503v2}D. A. Gomes, R.V. Maluf, C.A. S. Almeida, Annals of Physics 418, 168198 (2020).
\bibitem{s10052-020-7743-y}C. Ding, C. Liu, R. Casana, A. Cavalcante, Eur. Phys. J. C 80, 178 (2020).
\bibitem{1804.09911v2}A. Ovgun, K. Jusufi and I. Sakallı, Exact traversable wormhole solution in bumblebee gravity,Phys. Rev. D 99 (2019) 024042 [arXiv:1804.09911].
\bibitem{2407.13487v1}P. Sarmah, U.D. Goswami. Anisotropic cosmology in Bumblebee gravity theory[J]. arXiv preprint arXiv:2407.13487, 2024.
\bibitem{2411.18559v1}X.C. Zhu, R. Xu, and D.D. Xu. "Bumblebee Cosmology: Tests Using Distance- and Time-Redshift Probes." arXiv preprint arXiv:2411.18559v1 (2024).
\bibitem{2207.14423v2}D. Liang, R. Xu, X. Lu and L. Shao, Polarizations of Gravitational Waves in the Bumblebee Gravity Model,Phys. Rev. D 106 (2022) 124019 [arXiv:2207.14423].
\bibitem{S0370269324003435-main} K. M. Amarilo, et al. "Gravitational Waves Effects in a Lorentz–Violating Scenario." Physics Letters B 855 (2024): 138785.

\bibitem{Phys.Rev.D101(2).024040(2020}Z.H. Li, Ali Övgün. "Finite-distance gravitational deflection of massive particles by the Kerr-like black hole in the bumblebee gravity model." Physical Review D 101.2 (2020): 024040.
\bibitem{physrevd.9.2273}M. Kalb and P. Ramond, Classical direct interstring action, Phys. Rev. D 9 (1974) 2273.
\bibitem{119f92711c9024a90254f8741f4bf5081}W.F. Kao, W.B. Dai, S.Y. Wang, T.K. Chyi and S.Y. Lin, Induced Einstein-Kalb-Ramond theory and the black hole, Phys. Rev. D 53 (1996) 2244.
\bibitem{0210176v2}S. Kar, S. SenGupta, S. Sur. Static, spherically symmetric solutions, gravitational lensing, and perihelion precession in Einstein-Kalb-Ramond theory[J]. Physical Review D, 2003, 67(4): 044005.
\bibitem{1611.06936v2}S. Chakraborty and S. SenGupta, Strong gravitational lensing — a probe for extra dimensions and Kalb-Ramond field, JCAP 07 (2017) 045 [1611.06936].
\bibitem{2112.11945v2}K.K. Nair and A.M. Thomas, Kalb-Ramond field-induced cosmological bounce in generalized teleparallel gravity, Phys. Rev. D 105 (2022) 103505 [arXiv:2112.11945].
\bibitem{1207.3152v3}C.E. Fu, Y.X. Liu, K. Yang and S.-W. Wei,Q-form fields on p-branes, JHEP 10 (2012) 060[arXiv:1207.3152].
\bibitem{2308.06613}K. Yang, Y.Z. Chen, Z.Q. Duan, and J.Y. Zhao. "Static and Spherically Symmetric Black Holes in Gravity with a Background Kalb-Ramond Field." arXiv preprint arXiv:2308.06613v2 (2023).
\bibitem{2106.14602v1}H.M. Wang, and S.W. Wei. "Shadow Cast by Kerr-like Black Hole in the Presence of Plasma in Einstein-Bumblebee Gravity." arXiv preprint arXiv:2106.14602v1 (2021).
\bibitem{s10052-023-11231-5}W. Liu, X. Fang, J. Jing and J. Wang, QNMs of slowly rotating Einstein-Bumblebee black hole,Eur. Phys. J. C 83 (2023) 83 [arXiv:2211.03156].
\bibitem{s10052-024-13188}Z.Q. Duan, J.Y. Zhao, and K. Yang. "Electrically Charged Black Holes in Gravity with a Background Kalb–Ramond Field." European Physical Journal C 84 (2024).
\bibitem{2001.00460v2}R. Kumar, S.G. Ghosh, A. Wang. Gravitational deflection of light and shadow cast by rotating Kalb-Ramond black holes[J]. Physical Review D, 2020, 101(10): 104001.
\bibitem{2010.05298v1}L.A. Lessa, R. Oliveira, J.E.G. Silva and C.A.S. Almeida, Traversable wormhole solution with a background Kalb-Ramond field, Annals Phys. 433 (2021) 168604[arXiv:2010.05298]
\bibitem{s10052-022-10409-7}R.V. Maluf and C.R. Muniz, Exact solution for a traversable wormhole in a curvature-coupled antisym metric background field, Eur. Phys. J. C 82 (2022) 445 [2110.12202].
\bibitem{1911.10296} L.A. Lessa, et al. "Modified Black Hole Solution with a Background Kalb–Ramond Field." European Physical Journal C 80 (2020).
\bibitem{10052.10619}F. Atamurotov, D. Ortiqboev, A. Abdujabbarov, et al. Particle dynamics and gravitational weak lensing around black hole in the Kalb-Ramond gravity[J]. The European Physical Journal C, 2022, 82(8): 659.
\bibitem{2304.07761}A. Baruah, A. Övgün, A. Deshamukhya. Quasinormal modes and bounding greybody factors of GUP-corrected black holes in Kalb–Ramond gravity[J]. Annals of Physics, 2023, 455: 169393.
\bibitem{s10052-022-10619-z}F. Atamurotov, D. Ortiqboev, A. Abdujabbarov, et al. Particle dynamics and gravitational weak lensing around black hole in the Kalb-Ramond gravity[J]. The European Physical Journal C, 2022, 82(8): 659.
\bibitem{2304.10015} X.J. Wang, et al. "Rings and Images of Horndeski Hairy Black Hole Illuminated by Various Thin Accretions." arXiv preprint arXiv:2304.10015 (2023).
\bibitem{S0550321322003777}H.M. Wang, Z.C. Lin, S.W. Wei. Optical appearance of Einstein-Æther black hole surrounded by thin disk[J]. Nuclear Physics B, 2022, 985: 116026.
\bibitem{Image of a spherical black hole}J.P. Luminet, “Image of a spherical black hole with thin accretion disk,” Astron. Astrophys. 75 (1979) 228–235.
\bibitem{1304.5691}C. Bambi. Can the supermassive objects at the centers of galaxies be traversable wormholes? The first test format of strong gravity for mm/sub-mm very long baseline interferometry facilities[J]. Physical Review D, 2013, 87(10): 107501.
\bibitem{Falcke_2000_ApJ_528_L13}H. Falcke, and F. Melia, E. Agol. Astrophys. J. Lett. 528 (2000) L13.
\bibitem{Narayan_2019_ApJL_885_L33}R. Narayan, M. D. Johnson, and C. F. Gammie, Astrophys. J. Lett. 885 (2019) L33.
\bibitem{2212.12949}F. Atamurotov, M. Jamil, K. Jusufi. Quantum effects on the black hole shadow and deflection angle in the presence of plasma[J]. Chinese Physics C, 2023, 47(3): 035106.
\bibitem{1507.08545}G.S. Bisnovatyi-Kogan, O.Y. Tsupko. Gravitational lensing in plasmic medium[J]. Plasma Physics Reports, 2015, 41: 562-581.
\bibitem{1905.06615}G.S. Bisnovatyi-Kogan, and O.Y. Tsupko. "Gravitational Lensing in Presence of Plasma: Strong Lens Systems, Black Hole Lensing and Shadow." arXiv preprint arXiv:1905.06615 (2019).
\bibitem{1507.04217v2}V. Perlick, O.Y. Tsupko, and G.S. Bisnovatyi-Kogan, Phys. Rev. D 92, 104031 (2015), arXiv:1507.04217 [gr-qc]
\bibitem{1702.08768}V. Perlick, and O.Y. Tsupko. "Light Propagation in a Plasma on Kerr Spacetime: Separation of the Hamilton-Jacobi Equation and Calculation of the Shadow."arXiv:1702.08768 (2017).
\bibitem{1507.08131}F. Atamurotov, B. Ahmedov, A. Abdujabbarov. Optical properties of black holes in the presence of a plasma: The shadow[J]. Physical Review D, 2015, 92(8): 084005.
\bibitem{1807.06268}Y. Huang, Y.P. Dong, D.J. Liu. Revisiting the shadow of a black hole in the presence of a plasma[J]. International Journal of Modern Physics D, 2018, 27(12): 1850114.
\bibitem{lrr-2004-9}V. Perlick. Gravitational lensing from a spacetime perspective[J]. Living reviews in relativity, 2004, 7: 1-117.
\bibitem{1702.08768v2}V. Perlick and O.Y. Tsupko, Phys. Rev. D 95, 104003 (2017), arXiv:1702.08768 [gr-qc] .
\bibitem{Phys. Rev. D 81}B. Altschul, Q.G. Bailey, V.A. Kostelecký. Lorentz violation with an antisymmetric tensor[J]. Physical Review D—Particles, Fields, Gravitation, and Cosmology, 2010, 81(6): 065028.
\bibitem{0912.4852v1}B. Altschul, Q.G. Bailey, and V. A. Kostelecký. "Lorentz violation with an antisymmetric tensor." Physical Review D 80.12 (2009): 125008. arXiv:0912.4852v1 [gr-qc].
\bibitem{2412.00796}E.L.B. Junior, J.T.S.S. Junior, F.S.N. Lobo, et al. Periodical orbits and waveforms with spontaneous Lorentz symmetry-breaking in Kalb-Ramond gravity[J]. arXiv preprint arXiv:2412.00769, 2024.
\bibitem{2112.11227} Guo S, Li G R, Liang E W. Influence of accretion flow and magnetic charge on the observed shadows and rings of the Hayward black hole. Physical Review D, 2022, 105(2): 023024.
\bibitem{huyapeng} Gao X J, Sui T T, Zeng X X, et al. Investigating shadow images and rings of the charged Horndeski black hole illuminated by various thin accretions. The European Physical Journal C, 2023, 83(11): 1052.
\bibitem{2406.07300}S. Zare, et al. "Shadows, rings and optical appearance of a magnetically charged regular black hole illuminated by various accretion disks." arXiv preprint arXiv:2406.07300 (2024).
\bibitem{Afrin2022.93251}M. Afrin, and G.G. Sushant. "Testing Horndeski Gravity from EHT Observational Results for Rotating Black Holes." The Astrophysical Journal 932.51 (2022): 1-11.

\bibitem{2205.07787} Vagnozzi S, Roy R, Tsai Y D, et al. Horizon-scale tests of gravity theories and fundamental physics from the Event Horizon Telescope image of Sagittarius A*. Classical and Quantum Gravity, 2023, 40(16): 165007.

\bibitem{2205.07787.can22}M. Wielgus et al. (Event Horizon Telescope), Astrophys. J. Lett. 930, L19 (2022).
\bibitem{2205.07787.can18}K. Akiyama et al. (Event Horizon Telescope), Astrophys. J. Lett. 930, L15 (2022).
\bibitem{Irr 2013-1}M.A. Abramowicz, and P. C. Fragile. "Foundations of Black Hole Accretion Disk Theory." Living Reviews in Relativity 16 (2013).
\bibitem{2105.01173}Event Horizon Telescope Collaboration. First M87 event horizon telescope results. VIII. Magnetic field structure near the event horizon[J]. arXiv preprint arXiv:2105.01173, 2021.

\bibitem{2312.10678} Hoshimov H, Yunusov O, Atamurotov F, et al. Weak gravitational lensing and shadow of a GUP-modified Schwarzschild black hole in the presence of plasma. Physics of the Dark Universe, 2024, 43: 101392.
\bibitem{2206.04430} Zhang Z, Yan H, Guo M, et al. Shadows of Kerr black holes with a Gaussian-distributed plasma in the polar direction. Physical Review D, 2023, 107(2): 024027.
\bibitem{2209.01652}F. Atamurotov, et al. "Shadow and quasinormal modes of the Kerr–Newman–Kiselev–Letelier black hole." The European Physical Journal C 82.9 (2022): 831.
\bibitem{Eur. Phys. J. C 82 771 (2022}F. Sarikulov, F. Atamurotov, A. Abdujabbarov, et al. Shadow of the Kerr-like black hole[J]. The European Physical Journal C, 2022, 82(9): 771.
\bibitem{2106.07601}J. Bádia,, and F.E. Ernesto. "Shadow of Axisymmetric, Stationary and Asymptotically Flat Black Holes in the Presence of Plasma." Physical Review D 104.10 (2021): 104066.
\bibitem{2110.11704}A. Das, S. Ashis, and G. Sunandan. "Study of Circular Geodesics and Shadow of Rotating Charged Black Hole Surrounded by Perfect Fluid Dark Matter Immersed in Plasma." arXiv preprint arXiv:2110.11704 (2022).
\bibitem{Eur. Phys. J. C 82 659 (2022}S.Y. Hu, et al. "Observational signatures of Schwarzschild-MOG black holes in scalar-tensor-vector gravity: shadows and rings with different accretions." Eur. Phys. J. C 82 (2022): 885.
\bibitem{Eur. Phys. J. Plus 137 634 (2022}V.V. Dodonov, and A.V. Dodonov. "Magnetic moment invariant Gaussian states of a charged particle in a homogeneous magnetic field." The European Physical Journal Plus 137.5 (2022): 575.
\bibitem{Phys.Rev. D 104. 084015 (2021}F. Atamurotov, A. Ahmadjon, and W.B. Han. "Effect of plasma on gravitational lensing by a Schwarzschild black hole immersed in perfect fluid dark matter." Physical Review D 104.8 (2021): 084015.
\bibitem{2201.09879}W. Javed, I. Hussain, A. Övgün. Weak deflection angle of Kazakov–Solodukhin black hole in plasma medium using Gauss–Bonnet theorem and its greybody bonding[J]. The European Physical Journal Plus, 2022, 137(1): 1-14.
\bibitem{2207.06994}A. Das, A. Saha, S. Gangopadhyay. Shadow of Kottler black hole in the presence of plasma for a co-moving observer[J]. Classical and Quantum Gravity, 2022, 40(1): 015008.
\bibitem{hr}V. Perlick, O.Y. Tsupko, and G.S. Bisnovatyi-Kogan, In- fluence of a plasma on the shadow of a spherically sym-metric black hole, Phys. Rev. D 92, 104031 (2015).
\bibitem{2304.03660}E. Ghorani, B. Puliçe, F. Atamurotov, et al. Probing geometric proca in metric-palatini gravity with black hole shadow and photon motion[J]. The European Physical Journal C, 2023, 83(4): 318.
\bibitem{2211.04263}J. Yang, C. Zhang, Y. Ma. Shadow and stability of quantum-corrected black holes[J]. The European Physical Journal C, 2023, 83(7): 619.
\bibitem{1906.00873}S.E. Gralla, D.E. Holz, R.M. Wald. Black hole shadows, photon rings, and lensing rings[J]. Physical Review D, 2019, 100(2): 024018.
\bibitem{2008.00657}J. Peng, M. Guo, X.H. Feng. Influence of quantum correction on black hole shadows, photon rings, and lensing rings[J]. Chinese Physics C, 2021, 45(8): 085103.
\bibitem{2307.1236v2}Z.L. Wang. "Shadows and Rings of a de Sitter-Schwarzschild Black Hole." European Physical Journal Plus (2023). arXiv preprint arXiv:2307.12361.
\bibitem{2206.12820}S.J. Ma, T.C. Ma, J.B. Deng, and X.R. Hu. "Shadow of Schwarzschild Black Hole in the Cold Dark Matter Halo." arXiv preprint arXiv:2206.12820 (2023).
\bibitem{NTISCO} Bambi, C. (2017). Thin Accretion Disks. In: Black Holes: A Laboratory for Testing Strong Gravity. Springer, Singapore.
\bibitem{NT1} Novikov I D, Thorne K S. Astrophysics of black holes. Black holes (Les astres occlus), 1973, 1: 343-450.
\bibitem{NT2} D.N. Page, K.S. Thorne, Astrophys. J. 191, 499 (1974).

\bibitem{disk4} Li, Guo-Ping, and Ke-Jian He. Shadows and rings of the Kehagias-Sfetsos black hole surrounded by thin disk accretion. Journal of Cosmology and Astroparticle Physics 2021.06 (2021): 037.
\bibitem{disk5} Uniyal, Akhil, et al. Nonlinearly charged black holes: Shadow and Thin-accretion disk. New Astronomy 111 (2024): 102249.
\bibitem{disk6} Xiaogang L. Charged Black Holes in the Kalb-Ramond Background with Lorentz Violation: Null Geodesics and Optical Appearance of a Thin Accretion Disk. Chinese Physics C, 2025.

\bibitem{disk1} Falcke H, Melia F, Agol E. Viewing the Shadow of the Black Hole at the GalacticCenter. The Astrophysical Journal, 1999, 528(1): L13.
\bibitem{disk2} Johannsen T, Psaltis D. Testing the no-hair theorem with observations in the electromagnetic spectrum. II. Black hole images. The Astrophysical Journal, 2010, 718(1): 446.
\bibitem{disk3} Jaroszynski M, Kurpiewski A. Optics near Kerr black holes: spectra of advection dominated accretion flows. arXiv preprint astro-ph/9705044, 1997. 
\end{thebibliography}
\end{document}